\documentclass[prb,amsmath,amssymb]{revtex4}

\usepackage{graphicx}
\usepackage{dcolumn}
\usepackage{bm}

\begin{document}

\title{Telegraph-type versus
diffusion-type models of turbulent
relative dispersion}

\author{Kentaro Kanatani}
\author{Takeshi Ogasawara}
\author{Sadayoshi Toh}
\affiliation{Division of Physics and Astronomy, Graduate School of Science,
Kyoto University, Kyoto 606-8502, Japan}

\date{\today}

\begin{abstract}
Properties of two equations describing the evolution of 
the probability density function (PDF) of 
the relative dispersion in turbulent flow
are compared by investigating their solutions: 
the Richardson diffusion equation with the drift term 
and the self-similar telegraph equation 
derived by 
Ogasawara and Toh [J. Phys. Soc. Jpn. {\bf 75}, 083401 (2006)].
The solution of 
the self-similar telegraph equation
vanishes at a finite point, 
which represents persistent separation of a particle pair, 
while that of 
the Richardson equation
extends infinitely just after the initial time.
Each equation has a similarity solution, which 
is found to be an asymptotic solution 
of the initial value problem. 
The time lag has a dominant effect on the relaxation process 
into the similarity solution.
The approaching time to the similarity solution
can be reduced by 
advancing the time of the similarity solution appropriately.
Batchelor scaling, a scaling law
relevant to initial separation, is observed 
only for the telegraph case.
For both models, we estimate the Richardson constant,
based on their similarity solutions.
\end{abstract}

\maketitle

\section{Introduction}

The relative dispersion of particle pairs
is 
fundamental to 
the 
turbulent research
and has many practical applications to environmental and industrial problems,
such as the transport of pollutants in the atmosphere
and the fuel mixing in engines.
The study of turbulent relative dispersion has a long history
since the pioneering work by Richardson
\cite{richardson26:_atmos},
who observed the anomalous dispersion 
or superdiffusion of particle pairs 
and proposed the diffusion equation
describing the evolution of the probability density function (PDF)
for the pair separation.
Several attempts to modify the Richardson model
have been presented
\cite{batchelor52:_diffus_i,kraichnan66:_disper,
sawford01:_turbul}. 
Recent experiments and direct numerical simulations (DNSs)
showed that the Richardson model can well represent
the separation PDF
 in the inertial subrange
\cite{m.99:_richar_pair_disper_two_dimen_turbul,ott00,
boffetta02:_relat_disper_fully_devel_turbul,l.05:_lagran,
berg06:_backw}.
Nevertheless, the best model of 
the relative dispersion has not yet been
determined decisively.
This is because 
the inertial ranges
achieved both in 
experiments and DNSs up to now
are not wide enough 
to observe the full superdiffusive behavior.

The relative dispersion in turbulent flows has often been modeled
with the Reynolds-number dependence and the contribution from the three ranges
of turbulence, namely the energy-dissipative, 
the inertial and the energy-containing ranges
\cite{borgas04:_relat,franzese07,luthi07:_self}.
Indeed, the unified description of the relative dispersion is 
important especially for comparison of data
among models, experiments, 
and DNSs at moderate Reynolds number.
However, we limit our attention to
the relative dispersion in the infinitely-extended inertial range, 
namely in the case that Re $\rightarrow \infty$,
where the superdiffusive behavior can be observed,
and discuss the governing equation of the separation PDF there.
As is already mentioned, the Richardson model is supported
by several experiments and DNSs.
However, there are a few shortcomings in this model.
One of those of the Richardson model
is 
the absence of
the long-time correlation of the Lagrangian relative velocity
 of a particle pair
\cite{sawford01:_turbul,sokolov99:_two},
which exists in real turbulent flows.
This correlation gives rise to persistent separation of pairs, 
which should obey the self-similarity as shown in the recent DNS
\cite{boffetta02:_statis}.

In order to overcome this 
shortcoming, Ogasawara and Toh
\cite{ogasawara06:_model} 
devised another model of turbulent relative dispersion,
the self-similar telegraph model.
This model is derived taking into account
the persistency of pair separation
on the basis of Sokolov's picture
\cite{sokolov99:_two,i.00:_ballis_richar},
where the relative separation
consists of persistent expansion and compression.
As a result, the governing equation
of the PDF of relative separation 
 in this model
has a second-order time derivative term.
This is the reason why this model is called
'telegraph,' while in the other previous theories
such as the Richardson model the governing equation
of PDF is diffusion-type.
The term 'self-similar' comes from
scale-dependent coefficients 
reflecting the self-similarity of the dispersion process
in the inertial range.
Owing to this inclusion of the self-similarity,
solutions of the self-similar telegraph equation do not approach
those of its corresponding diffusion equation,
whereas those of the usual telegraph equation appearing in the problems
of molecular diffusion or Brownian motion quickly relax into
those of the diffusion equation.
This indicates that the self-similar telegraph model has
essentially distinct properties from those of its diffusion-type counterpart.

In this paper, we investigate 
time-integrated solutions
of the self-similar telegraph equation 
as well as 
those of its diffusion-type counterpart.
In the literature, the realizability of the similarity solution
of the Richardson diffusion equation 
has implicitly been assumed,
and thereby the agreements between the similarity solution and
the separation PDF obtained by experiments or DNSs
have been discussed. 
However, the similarity solution
becomes a delta function 
at the origin in the limit $t \rightarrow 0$.
This requires that a particle pair be located
at the same place at $t = 0$.
This condition is difficult to set in 
experiments and DNSs.
In addition, the viscous effects in
the dissipation range may also contaminate
the pure inertial-range behavior
 of the dispersion process.
For this reason, the initial separation of a pair is 
finite in 
experiments and DNS,
and hence the initial condition with a nonzero separation is desired.

Furthermore, the recent experiment\cite{n.06} suggested that 
the separation PDF obtained by their experiment agreed well with 
the similarity solution of Batchelor's diffusion equation 
rather than Richardson's mentioned above.
However, their claim is contradictory in that 
the similarity solution of the Batchelor diffusion equation implies 
the usual Richardson scaling law for the mean-square separation 
$\langle r^2 \rangle \propto t^3$, 
whereas they found the (modified) Batchelor scaling law 
instead of it, as was also reported in Ref.~\onlinecite{m.06}.
These results stimulate us to investigate 
the short time behavior of the solution of 
the Richardson diffusion equation with finite initial separation.
To our knowledge, this issue has never been treated in the literature.

Motivated by the above,
we numerically solve the governing equations of the PDF
under appropriate initial conditions with finite separations,
and observe the behaviors of the time-integrated solutions.
Comparisons of the solutions between
the telegraph-type and the diffusion-type model
are also made to reveal the characteristics of the models.

The remainder of the paper is organized as follows.
In Sec.~\ref{sec:telegraph_Palm} we introduce 
the self-similar telegraph
equation together with its diffusion-type counterpart.
Section~\ref{sec:simulations} 
explains the settings of the simulation.
The results of the simulation is presented
in Sec.~\ref{sec:results}, where the 
Richardson constants
are also obtained and compared for the two models.
In Appendix~\ref{sec:decrement},
the decrement of the total probability
is examined for the self-similar telegraph model,
because the conservation of it is not guaranteed
for this model.

\section{\label{sec:telegraph_Palm}Self-similar telegraph
and Palm equations}

In this section, we introduce the self-similar telegraph model of turbulent relative dispersion.
In this model, the evolution of the spherically 
symmetric PDF of pair separation $P(r,t)$
in $d$-dimensional isotropic turbulence is described by the following
self-similar telegraph equation
\cite{ogasawara06:_model}
\begin{equation}
  \frac{T_c(r)}{\lambda} \frac{\partial^2 P}{\partial t^2}
  + \frac{\partial P}{\partial t} = \frac{\partial}{\partial r}
  \left[D(r) r^{d-1} \frac{\partial}{\partial r} 
    \left(\frac{P}{r^{d-1}}\right)\right]
  + \sigma \frac{\partial}{\partial r}[v(r)P],
\label{eq:telegraph}
\end{equation}
where $T_c(r)$, $D(r)$ and $v(r)$ are a characteristic 
time scale, a diffusion coefficient 
and the Lagrangian relative velocity, respectively,
of a particle pair of separation $r$.
The parameters $\lambda$ and $\sigma$ characterize
the turbulent field considered.
The coefficients $T_c(r)$, $D(r)$ and $v(r)$ are
assumed to obey the following scaling laws:
\begin{subequations}
\label{eq:scalings}
\begin{eqnarray}
  T_c(r) &=& \check{A}^{-1} r^s, \label{eq:time} \\
  D(r)   &=& \check{A} \lambda^{-1} r^{2-s}, \\
  v(r)   &=& \check{A} r^{1-s}. \label{eq:velocity}
\end{eqnarray}
\end{subequations}
Here, $\check{A}$ is a dimensional constant and
$s$ a scaling exponent: 
$s = 2/3$ for Kolmogorov scaling
and $s = 2/5$ for Bolgiano-Obukhov scaling.
The last term on the r.h.s. of 
Eq.~(\ref{eq:telegraph}) is
a drift term, with the drift velocity being $- \sigma v(r)$.
$\lambda^{-1}$ represents the persistency of the separation,
and corresponds to the persistent parameter
introduced by Sokolov
\cite{sokolov99:_two}.

In the derivation of Eq.~(\ref{eq:telegraph})
\cite{ogasawara06:_model},
a parameter $\delta$ was also used,
and has a relation with $\lambda$ and $\sigma$
via $\lambda \sigma = d - 2s + \delta$.
The physical meaning of $\delta$ is
the difference between the two transition rates
of the direction of the separation,
from expansion to compression and from compression
to expansion.
Hereafter, we use $\delta$ as 
a control parameter
instead of $\sigma$, following the previous papers
\cite{ogasawara06:_model,ogasawara06:_turbul}.

If the effects of the finite separation
and the finite correlation of the relative velocity are not considered,
the first term on the l.h.s. of Eq.~(\ref{eq:telegraph})
is omitted, and then Eq.~(\ref{eq:telegraph}) can be reduced to the following
diffusion equation:
\begin{equation}
  \frac{\partial P}{\partial t} = \frac{\partial}{\partial r}
  \left[D(r) r^{d-1} \frac{\partial}{\partial r} 
    \left(\frac{P}{r^{d-1}}\right)\right]
  + \sigma \frac{\partial}{\partial r}[v(r)P].
\label{eq:Palm}
\end{equation}
This equation has the same form as the Richardson
diffusion equation with the drift term.
If $\sigma = 0$, i.e. $\delta = 2s - d$,
Eq.~(\ref{eq:Palm}) is the Richardson diffusion
equation itself.
The addition of the drift term to 
the Richardson equation was discussed by Palm in 1957
\cite{monin75:_statis_fluid_mechan}.
Recently, Goto and Vassilicos
\cite{goto04:_partic}
derived the same equation as Eq.~(\ref{eq:Palm}).
We refer to Eq.~(\ref{eq:Palm}) as Palm equation.

Before we perform the time integration of 
Eqs.~(\ref{eq:telegraph}) and (\ref{eq:Palm}),
the nondimensionalization is made for convenience.
We normalize space by the nonzero initial relative separation
 $R$ $(\neq 0)$ 
of particle pairs
and time by the corresponding timescale $T_c(R)$.
Consequently, new dimensionless space and time variables are defined as
$\tilde{r} = r / R$ and $\tilde{t} = t / T_c(R) = \check{A} t / R^s$ 
respectively.
We also introduce the normalized separation PDF 
$\tilde{P}(\tilde{r},\tilde{t}) = R P(r,t)$.
Substituting Eqs.~(\ref{eq:scalings}) into 
Eqs.~(\ref{eq:telegraph}) and (\ref{eq:Palm}), 
eliminating the dimensional quantities
and dropping the superposed tildes from the dimensionless quantities
give the following nondimensionalized versions of the equations:
\begin{eqnarray}
  r^s \frac{\partial^2 P}{\partial t^2}
  + \lambda \frac{\partial P}{\partial t}
  = \frac{\partial}{\partial r}
  \left[r^{1-s+d} \frac{\partial}{\partial r} 
    \left(\frac{P}{r^{d-1}}\right)\right]  
  + (d - 2s + \delta) \frac{\partial}{\partial r}(r^{1-s}P),
\label{eq:telegraph2}
\end{eqnarray}
\begin{eqnarray}
  \lambda \frac{\partial P}{\partial t}
  = \frac{\partial}{\partial r}
  \left[r^{1-s+d} \frac{\partial}{\partial r} 
    \left(\frac{P}{r^{d-1}}\right)\right]  
  + (d - 2s + \delta) \frac{\partial}{\partial r}(r^{1-s}P).
\label{eq:Palm2}
\end{eqnarray}
In the following, 
we treat Eqs.~(\ref{eq:telegraph2}) and (\ref{eq:Palm2}) 
instead of Eqs.~(\ref{eq:telegraph}) and (\ref{eq:Palm}).

\section{\label{sec:simulations}Simulations}

\subsection{\label{subsec:parameters}Parameters}

We set the values of the parameters 
$d$ and $s$ to 2 and $2/5$, respectively, assuming the case of
two-dimensional free convection (2DFC) turbulence,
which is characterized by the Bolgiano-Obukhov scaling
\cite{monin75:_statis_fluid_mechan,toh94:_entrop,
bistagnino07:_lagran_bolgian}.
This is because 
we have recently estimated the values of the control parameters, 
$\lambda$ and $\delta$, using exit time statistics
\cite{ogasawara06:_turbul}.
For $\lambda$ and $\delta$ 
we took the values 
$5.2$ and $-0.77$, respectively, 
from Ref.~\onlinecite{ogasawara06:_turbul}.
Since one of the main goals is to compare the properties 
of the different models of turbulent relative dispersion,
this choice of the parameters is sufficient 
unless the variation of the parameters
yields qualitative changes of the solutions of the model equations.
In addition, it is worth noting that
the one-particle dispersion in the 2DFC turbulence
was also investigated 
only recently
\cite{bistagnino07:_lagran_bolgian}.

\subsection{\label{sec:conditions}Initial and boundary conditions}

We aim to examine the case that the pairs of the particles
have the same nonzero separation at the initial time, 
and therefore, the initial condition 
should be represented by a 
Dirac's delta function
located at a certain finite point. To mimic this initial condition,
  we employed 
the following finite-width
PDF as 
an alternative initial condition: 
\begin{equation}
	P(r,0) =
	\left\{
		\begin{array}{cc}
			\displaystyle
			\frac{1}{2w}
			\left(1 + \cos\left(\pi \frac{r - 1}{w}
			\right) \right) & |r - 1| < w \\
			0 & |r - 1| > w,
		\end{array}
	\right.
\label{eq:initial}
\end{equation}
where $w$ 
represents the width 
of the initial PDF.
As is mentioned in the last part of Sec.~\ref{sec:telegraph_Palm}, 
the spatial scale is normalized
by the initial separation, so that the initial PDF (\ref{eq:initial})
is distributed around $1$.
Hereafter, we regard $1$ as the typical initial relative separation.
Note that this value cannot be strictly referred to as the mean value
since the PDF should be spherically integrated 
to obtain the mean value of relative separation
in two or three dimension.
In the following, 
we consider the two cases, $w = 1$ and $0.1$,
to see the effects of the width of the initial PDF.
Narrower widths than $0.1$ cause the numerical instability
of the scheme for the present interval of grids and steeper ridges
at the bounds of the time-integrated solution 
of the self-similar telegraph equation
(see Subsec.~\ref{sec:short} and Fig.~\ref{fig:PDFs}).

For the self-similar telegraph equation (\ref{eq:telegraph2}),
another initial condition $[\partial P(r,t) / \partial t]_{t=0} = 0$
is required. 
This condition implies 
the symmetry between the extending and compressing pairs
at the initial time,
which reflects the situation where
the initial placement of 
pairs is uncorrelated with the velocity field
of 
turbulent flow.
Finally, we set the boundary conditions to
$P(0,t) = P(\infty,t) = 0$.

\subsection{Numerical method}

Time integration of Eqs.~(\ref{eq:telegraph2}) and (\ref{eq:Palm2}) 
is performed with the Crank-Nicolson scheme,
which is second order accurate in time,
after the transformation of the spatial variable.
We define a new spatial coordinate $r'$ as 
$r^s/s$ and 
$r/(r + L)$ for 
 Eqs.~(\ref{eq:telegraph2}) and (\ref{eq:Palm2}), respectively.
Here $L$ is an arbitrary parameter,
which determines "local density" of grids in $r$ 
space.
The former transformation leads to the constant
Courant number 
$c = r^{1-s} (\Delta t / \Delta r) = \Delta t / \Delta r'$ 
with uniform grids in $r'$ 
space, while
the latter transforms $r \in [0,\infty)$ to the 
finite computational plane $r' \in [0,1]$. 

For the self-similar telegraph model, grids are added
above the maximum grid of the computational plane
and remove the grids of the same number as 
the added ones when the maximum separation
of the PDF nearly reaches the maximum grid.
In this way, the number of the grids is always 
conserved during the computation
(although the interval between the grids
are increased at every change in the grids).
The grids of $82,000$ and $100,000$ were used
to compute the solutions of Eqs.~(\ref{eq:telegraph2}) and (\ref{eq:Palm2}), 
respectively.
For the self-similar telegraph model, 
we set the initial grid interval $\Delta r'_{init}$ to $10^{-4}$, 
while for the Palm model,
we set $L$ to $0.2$ or $10$ for the short-time 
and $100$ or $1000$ for the long-time behavior.

\section{\label{sec:results}Results and discussion}

\subsection{\label{sec:short}Short-time behavior}

\begin{figure}
\begin{tabular}{cc}
\scalebox{0.6}{
\includegraphics{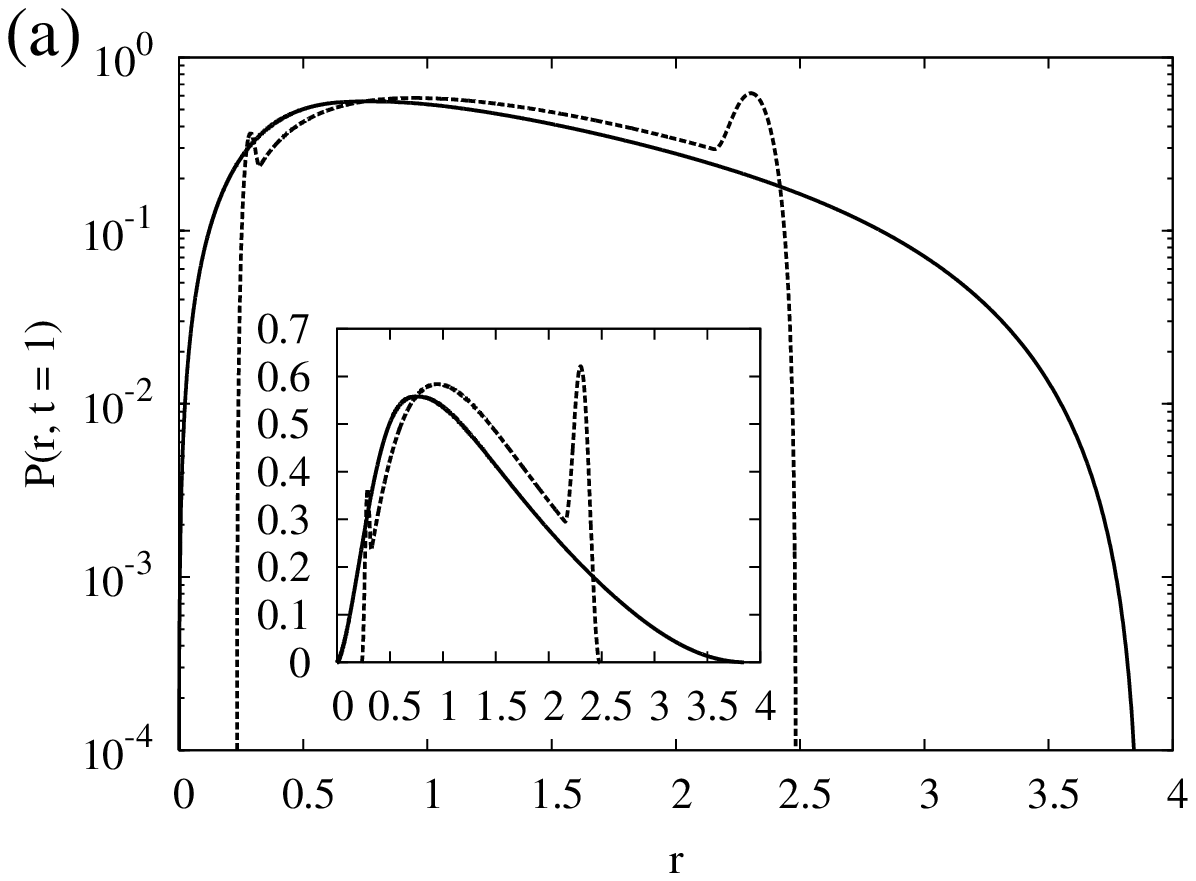}
} &
\scalebox{0.6}{
\includegraphics{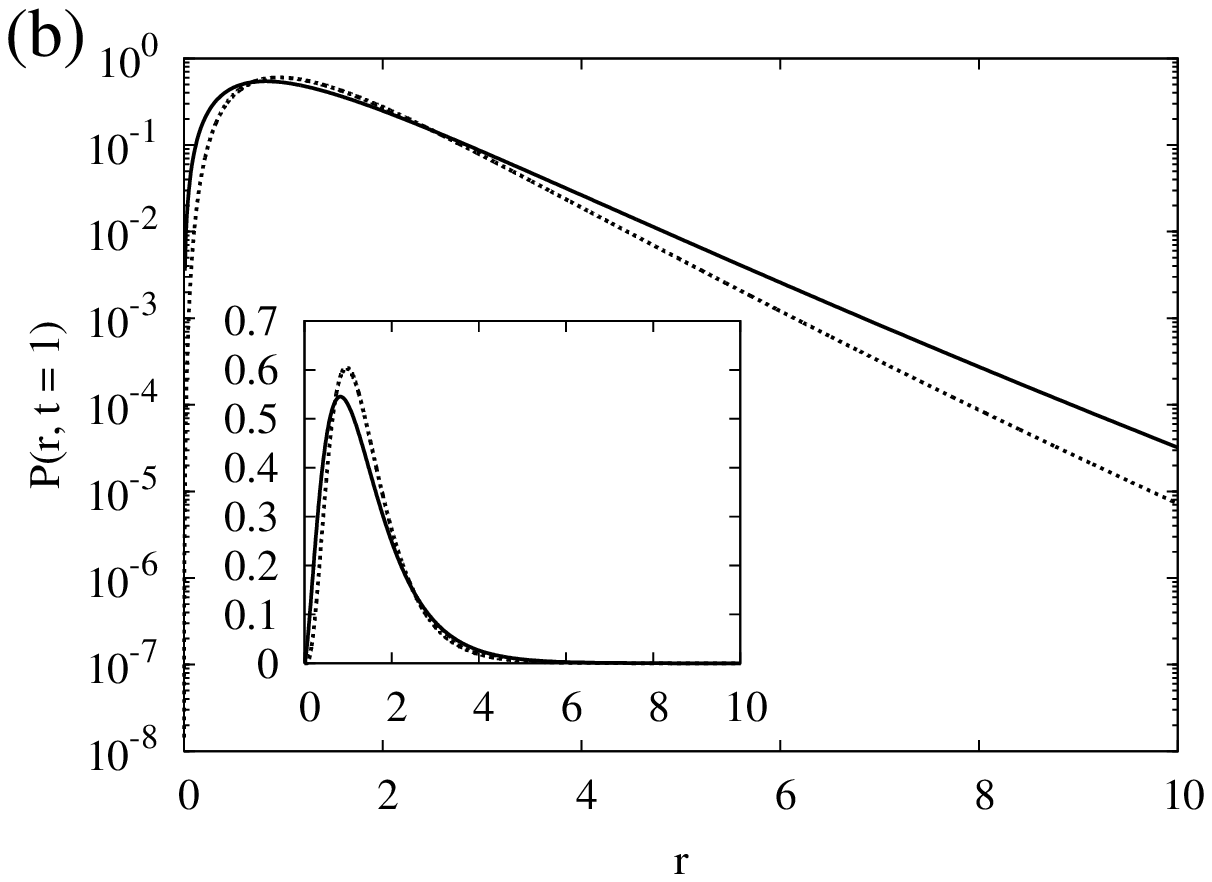}
} \\
\end{tabular}
\scalebox{0.6}{
\includegraphics{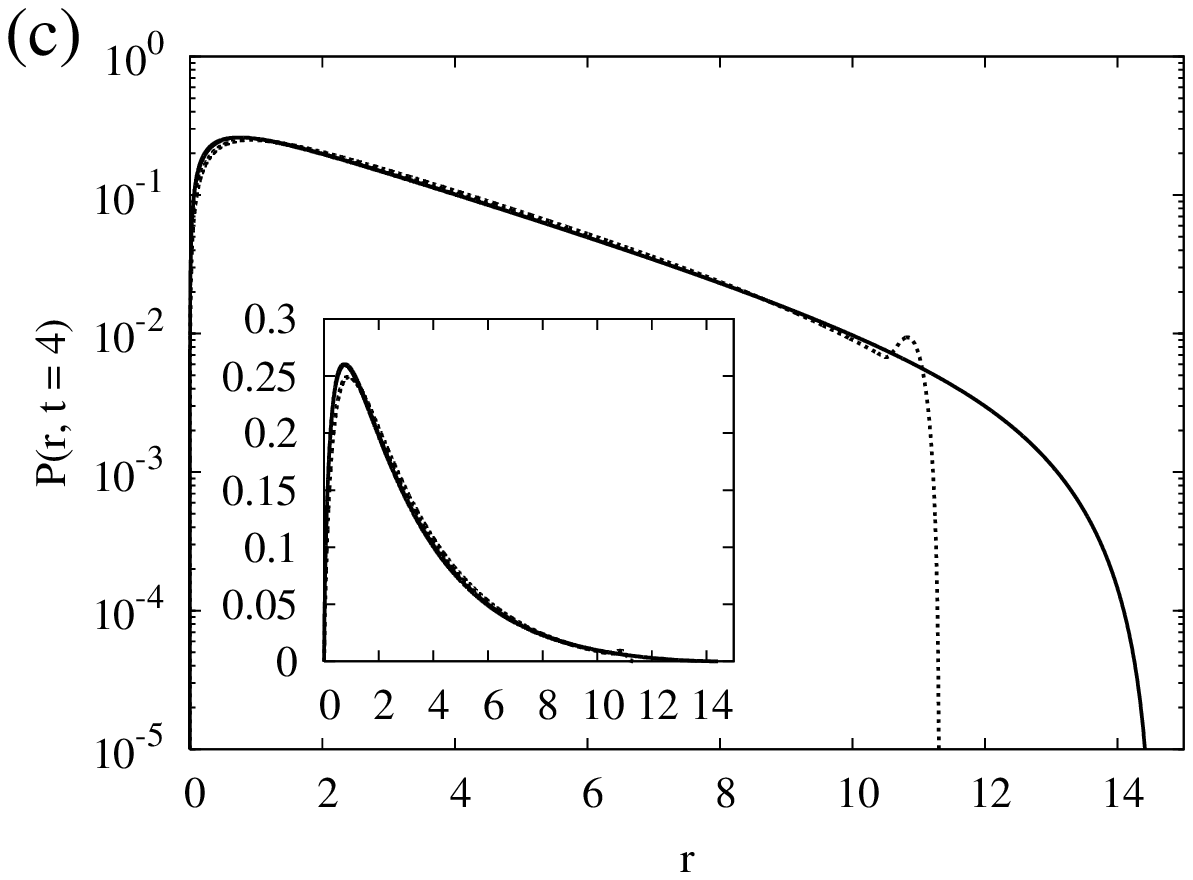}
} 
\caption{\label{fig:PDFs}
Time-integrated solutions of the self-similar telegraph equation~(\ref{eq:telegraph2}) and of the Palm equation~(\ref{eq:Palm2}). The solid lines denote the case of $w = 1$, while the dashed lines $w = 0.1$. (a) $t = 1$ for the telegraph case, (b) $t = 1$ for the Palm case and (c) $t = 4$ for the telegraph case. The insets are the same plots in linear scale.
}
\end{figure}

The 
time-integrated solutions of 
Eqs.~(\ref{eq:telegraph2}) and (\ref{eq:Palm2}) are shown
in Fig.~\ref{fig:PDFs} 
at different times.
While the solution of Eq.~(\ref{eq:Palm2})
extends infinitely at finite time,
that of Eq.~(\ref{eq:telegraph2}) 
bounds at a finite point.
This point corresponds to the maximum separation of 
pairs, which is exactly obtained by 
direct integration of 
the equation
$dr/dt = v(r)$ with
the scaling law of the velocity difference
 (\ref{eq:velocity})
to give in the nondimensionalized form
\begin{equation}
	r_{max}(t) = (r_{max}^s(0) + s t)^{1/s}.
\label{eq:separation}
\end{equation}
The existence of the maximum point in 
the solution is a manifestation of the finite separation of a pair,
which is included in the self-similar telegraph model.

For the case of $w = 0.1$ 
(Figs.~\ref{fig:PDFs}(a) and \ref{fig:PDFs}(c))
of the self-similar telegraph model,
we can observe ridges 
at the edges of the PDF.
The similar behavior has been found in the model of 
Ref.~\onlinecite{sokolov99:_drude}, 
based on a L\'evy-walk stochastic approach.
The PDF predicted by their model has peaks at the side edges.
Since our approach is similar to theirs,
the reason for the presence of the ridges should be consistent:
the distribution of the relative velocity is not taken into account.
Particle pairs having never been compressed or expanded from the initial time 
should be accumulated at these advancing or receding ridges, respectively.
To see this, let $r_{max}(0)$ in Eq.~(\ref{eq:separation}) be equal to 1,
the center of the initial PDF.
Then, $r_{max}(1) \simeq 2.32$, consistent with the position of 
the advancing ridge in Fig.~\ref{fig:PDFs}(a).
However, these ridges get smaller 
as time elapses owing to the effect of diffusion, 
i.e. the random changes of the direction of relative velocity.
In Fig.~\ref{fig:PDFs}(c),
at $t = 4$, the receding ridge disappears
and only the remnant of
the reduced advancing 
ridge can be observed
in the case of $w = 0.1$.

\subsection{Long-time behavior}

\begin{figure}
\scalebox{0.6}{
\includegraphics{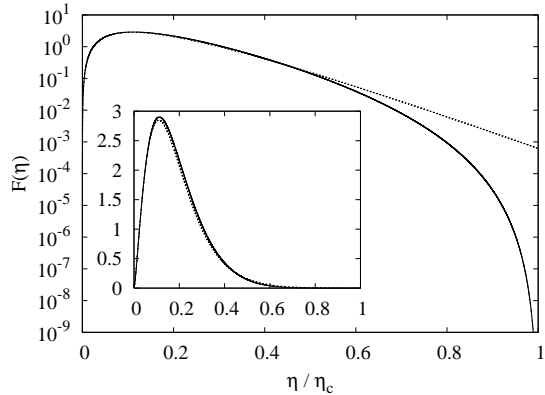}
}
\caption{\label{fig:solutions} 
Similarity solutions of the self-similar telegraph
equation~(\ref{eq:telegraph2}) (solid line) and the Palm
equation~(\ref{eq:Palm2}) (dashed line) in the similarity form $F(\eta) = (t / \lambda)^{1/s} P_s(r,t)$. The similarity variable $\eta$ is normalized by $\eta_c \equiv s \lambda$, corresponding to the maximum separation for the self-similar telegraph equation (\ref{eq:telegraph2}). $s = 2 / 5$, $d = 2$, $\lambda = 5.2$ and $\delta = -0.77$. The inset is a linear plot of the same figure. 
}
\end{figure}

Equation (\ref{eq:telegraph2}) can be reduced to
a second-order ordinary differential equation
with the similarity variable 
$\eta = \lambda r^s / t$.
Therefore it has a similarity solution,
and so for Eq.~(\ref{eq:Palm2}).
Although the similarity solution of Eq.~(\ref{eq:telegraph2})
has no analytical form, that of Eq.~(\ref{eq:Palm2}) does
and reads
\begin{equation}
	P_s(r,t) = 
        \frac{C}{t^{1/s}} \left(\frac{\lambda r^s}
	{s^2 t} \right)^{(2s - \delta - 1)/s}
	\exp \left(-\frac{\lambda r^s}{s^2 t} \right),
\label{eq:similarity}
\end{equation}
where
$C$ is the dimensionless normalization factor
\begin{equation}
  C = s \left(\frac{\lambda}{s^2 
    } \right)^{1/s}
	\bigg/ \Gamma \left(\frac{2s - \delta}{s} \right),
\end{equation}
and $\Gamma(x) = \int^{\infty}_0 e^{-t} t^{x-1} dt$ is the gamma function.
The subscript $s$ of $P_s$ denotes the similarity solution.
In Fig.~\ref{fig:solutions}, we show 
the similarity solutions of Eqs.~(\ref{eq:telegraph2}) and (\ref{eq:Palm2})
in the similarity form
$F(\eta) = (t / \lambda)^{1/s} P_s(r,t)$,
with the values of the parameters 
specified in Subsec.~\ref{subsec:parameters}.

In contrast to the solution of the Palm equation,
that of the self-similar telegraph equation
has an upper 
bound in space, and vanishes for
$\eta > \eta_c \equiv s \lambda$,
which is attributed to the singular point
of the reduced ordinary differential equation
\cite{ogasawara06:_model}.
This yields the maximum separation 
$r_{max}(t) = (s t)^{1/s}$,
which is also obtained by neglecting 
$r_{max}(0)$
in Eq.~(\ref{eq:separation}).

\begin{figure}
\begin{tabular}{cc}
\scalebox{0.6}{
\includegraphics{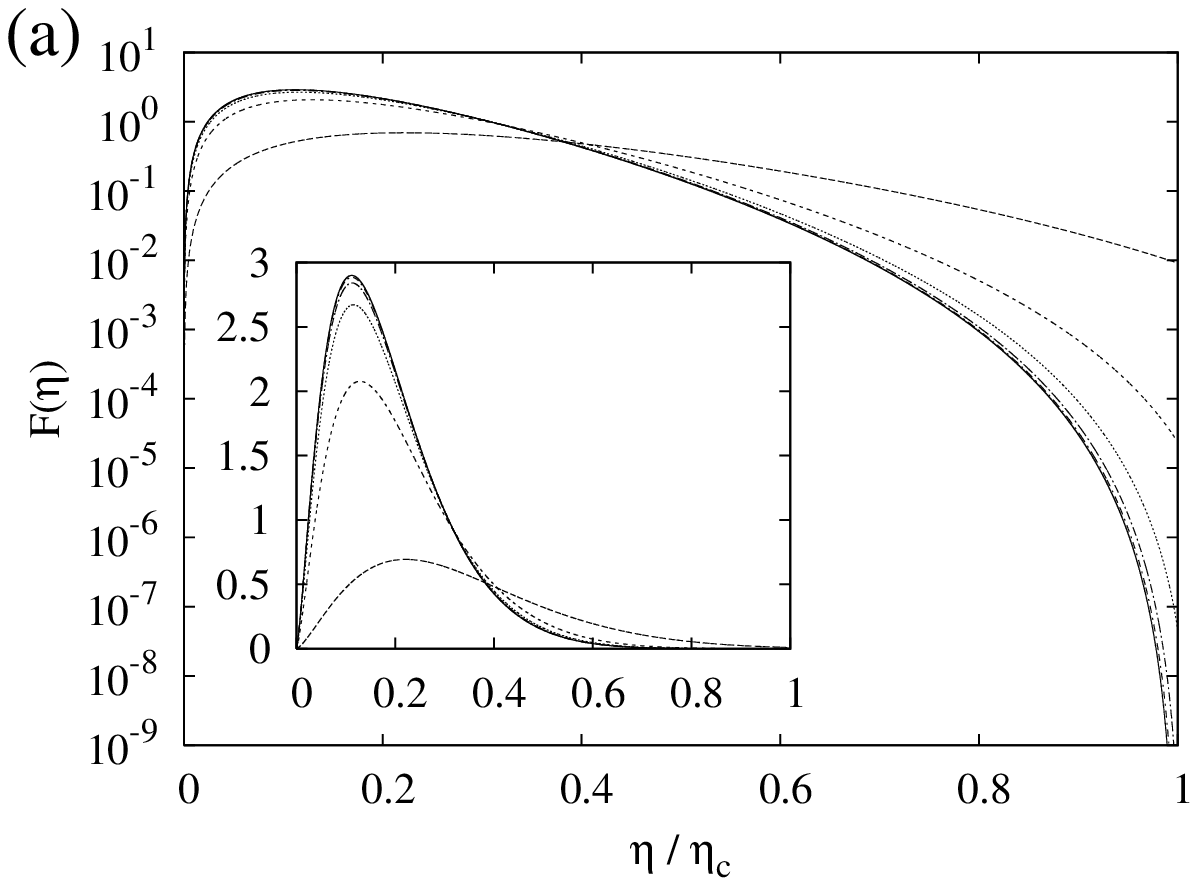}
} &
\scalebox{0.6}{
\includegraphics{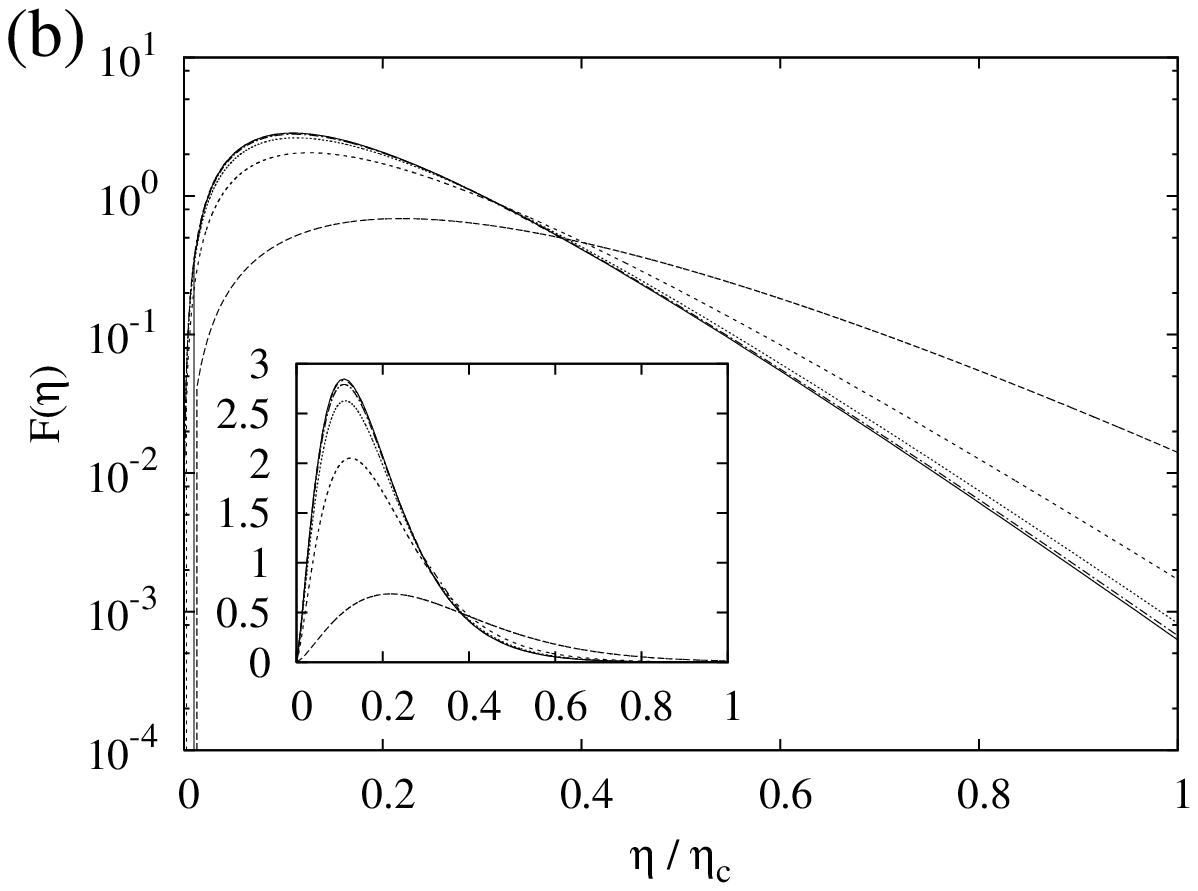}
}
\end{tabular}
\caption{\label{fig:approach}
Time-integrated solutions of (a) the self-similar telegraph
equation~(\ref{eq:telegraph2}) and (b) the Palm
equation~(\ref{eq:Palm2}) in the similarity form $F(\eta) = (t / \lambda)^{1/s} P(r,t)$, under the initial condition (\ref{eq:initial}) with $w = 1$, at time $t = 12$, $60$, $252$, $1020$ and $4092$ (from bottom to top at the maximum point), along with each similarity solution (solid line). The insets are linear plots of the same figures. 
}
\end{figure}

The similarity solutions tend to a delta function 
at the origin in the limit $t \rightarrow 0$, i.e. $P_s(r,0) = \delta(r)$,
as is expected by Eq.~(\ref{eq:similarity}) and
the fact that $r_{max}(0) = 0$ for the similarity solution
of the self-similar telegraph equation.
This initial condition is different from that adopted in the present paper.
However, we 
inferred that the effect of
the initial separation on the PDF is negligible
after a long time, and thus the solutions of
Eqs.~(\ref{eq:telegraph2}) and (\ref{eq:Palm2})
starting from our initial condition (\ref{eq:initial})
approach the corresponding similarity 
solutions.
In order to confirm this,
we further 
advanced the time-integrated solutions.
The resulting 
PDFs are transformed 
into the similarity form
$F(\eta)$
and are plotted in Fig.~\ref{fig:approach}.
From this figure,
we found that the solutions 
with the initial condition (\ref{eq:initial})
approach the similarity ones as time elapses,
and finally almost reach them at $t = 4092$,
so that we cannot distinguish them.
This suggests that the similarity solutions are
asymptotic solutions of the initial value problem.
The same behaviors as that shown in Fig.~\ref{fig:approach}
can be obtained for $w = 0.1$.

In order to analyze this behavior quantitatively, we 
introduce the root-squared difference between the 
time-integrated and the similarity PDF as
\begin{equation}
        \| \delta P \| = 
        \sqrt{\int^{\infty}_0 \bigl|P(r,t) - P_s(r,t)\bigr|^2 
          dr}.
\label{eq:norm}
\end{equation}
This quantity vanishes if the separation PDF
has the same form as that of the similarity solution.
Thus, we can 
quantify the degree of the 
proximity of the two solutions by calculating this quantity.

\begin{figure}
\scalebox{0.6}{
\includegraphics{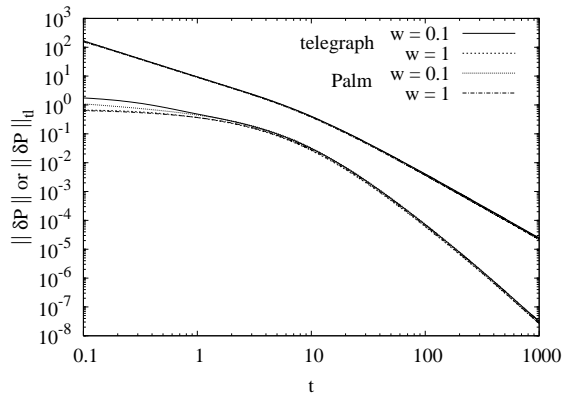}
}
\caption{\label{fig:diffs}
Temporal evolution of the root-squared differences between the time-integrated and the similarity solutions, $\| \delta P \|$ or $\| \delta P \|_{tl}$, defined in (\ref{eq:norm}) or (\ref{eq:norm2}), for both equations and widths. The upper curves denote $\| \delta P \|$, while the lower $\| \delta P \|_{tl}$: advancing times $t_{at}$ in Eq.~(\ref{eq:norm2}) are set to $8.5$, $8.3$, $8.3$ and $8.2$ for the telegraph with $w = 0.1$, the telegraph with $w = 1$, the Palm with $w = 0.1$ and the Palm with $w = 1$, respectively.
}
\end{figure}

In Fig.~\ref{fig:diffs},
we plotted the temporal 
evolution of 
$\| \delta P \|$ for 
both models and 
widths.
From this figure, the scaling laws 
$\| \delta P \| \propto t^{-\beta}$ 
can be found in both short-time and  long-time regimes with 
$\beta \simeq 1.1$ and $2.2$, respectively.
The transition between the two regimes occurs
around $t = 10$.
No qualitative difference between the two models is found.
The difference between the two 
values of $w$ in the initial condition (\ref{eq:initial}) can scarcely be seen 
throughout the whole time scales shown in the figures.
In this case, the relaxation 
processes into the similarity 
solutions 
appear to be rather universal, 
independent of 
both models and widths.
However, 
as is illustrated below, these apparent universal behaviors
are mainly due to the effect of time lag.

As an explanation of the origin of the above scaling behaviors,
we consider the effect of time lag,
because the similarity solutions are localized at the origin
in the limit $t \rightarrow 0$,
whereas the initial condition for the time-integrated
solutions (\ref{eq:initial}) implies nonzero initial separations.
We assume that the initial condition (\ref{eq:initial}) can be 
approximately described by the similarity solution at a certain time $t_0$.
Then, its subsequent time evolution is the same 
as the original similarity solution after $t_0$ only for the Palm case,
since for the telegraph case it depends on the additional condition 
$[\partial P(r,t) / \partial t]_{t=0}$.
We can now replace $P(r,t)$ by $P_s(r,t+t_0)$ in the right hand side of 
Eq.~(\ref{eq:norm}), and substitute 
Eq.~(\ref{eq:similarity}) to obtain
\begin{eqnarray}
  \lefteqn{\int^{\infty}_0 \bigl|P_s(r,t + t_0) - P_s(r,t)\bigr|^2 dr}
  \nonumber \\
  && = \frac{C^2}{2^{\gamma} s} 
  \left(\frac{s^2}{\lambda t}\right)^{1/s}
  \Gamma(\gamma)
  \left\{1 + \left(\frac{t}{t + t_0}\right)^{1/s}
    -2 \left(\frac{2 t}{2 t + t_0}\right)^{\gamma}
    \left(\frac{t + t_0}{t}\right)^{(2s - \delta - 1)/s}\right\},
\label{eq:Palm_diff}
\end{eqnarray}
where
\begin{equation}
  \gamma = \frac{4s - 2\delta - 1}{s}.
\end{equation}
We examine the two limiting cases for Eq.~(\ref{eq:Palm_diff}),
namely the long-time and the short-time limits.
The calculation for the former case is straightforward
and gives
\begin{eqnarray}
  \int^{\infty}_0 \bigl|P_s(r,t + t_0) - P_s(r,t)\bigr|^2 dr
  \simeq \frac{C^2}{2^{\gamma + 2} s^3}
  \left(\frac{s^2}{\lambda t}\right)^{1/s}
  \Gamma(\gamma)
  (\gamma s^2 + 1) \left(\frac{t_0}{t}\right)^2 
  \quad
  (t_0 \ll t).
\label{eq:long_time}
\end{eqnarray}
For the latter case, it is obvious that
\begin{eqnarray}
  \int^{\infty}_0 \bigl|P_s(r,t + t_0) - P_s(r,t)\bigr|^2 dr
  \simeq \int^{\infty}_0 P^2_s(r,t) dr
  = \frac{C^2}{2^{\gamma} s} 
  \left(\frac{s^2}{\lambda t}\right)^{1/s}
  \Gamma(\gamma) 
  \quad
  (t_0 \gg t),
\label{eq:short_time}
\end{eqnarray}
since 
only the $P^2_s(r,t)$ term in the integrand of Eq.~(\ref{eq:Palm_diff})
diverges in the limit $t \rightarrow 0$. 
Equations~(\ref{eq:long_time}) and (\ref{eq:short_time}) lead to
the scaling laws $t^{-1-1/(2s)} = t^{-2.25}$ and 
$t^{-1/(2s)} = t^{-1.25}$,
respectively, for the root-squared difference.
This explains the above scaling behaviors of $\| \delta P \|$ 
in Fig.\ref{fig:diffs}.

\begin{figure}
\scalebox{0.6}{
\includegraphics{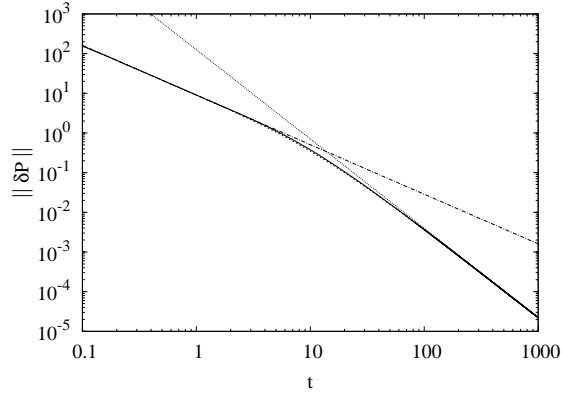}
}
\caption{\label{fig:fitting}
Fitting of $\| \delta P \|$ with the root of Eq.~(\ref{eq:Palm_diff}), in the case of the Palm model (\ref{eq:Palm2}) with $w = 1$. The solid and the dashed lines represent $\| \delta P \|$ and the root of Eq.~(\ref{eq:Palm_diff}) with the fitting parameter $t_0 \simeq 8.2$, respectively. The dotted and the dot-dashed lines are the roots of Eqs.~(\ref{eq:long_time}) and (\ref{eq:short_time}), respectively, with $t_0 \simeq 8.2$.
}
\end{figure}

Figure~\ref{fig:fitting} displays the result of the fitting of
$\| \delta P \|$ 
in the case of the Palm model (\ref{eq:Palm2}) 
with $w = 1$,
shown in Fig.~\ref{fig:diffs},
with the square root of Eq.~(\ref{eq:Palm_diff}).
Here, the time lag $t_0$ 
of value about $8.2$ for the square root of Eq.~(\ref{eq:Palm_diff})
gives the best fit of $\| \delta P \|$.
The agreement is excellent.
This result ensures our expectation that the time lag has the dominant effect
on the decay of $\| \delta P \|$ in time.

By removing the effect of the time lag,
more rapid approach of the time-integrated solution
to the similarity solution can be realized.
In Fig.~\ref{fig:diffs}, we also plotted the following quantity:
\begin{equation}
        \| \delta P \|_{tl} = 
        \sqrt{\int^{\infty}_0 \bigl|P_s(r,t + t_{at}) - P(r,t)\bigr|^2 dr}.
\label{eq:norm2}
\end{equation}
Here, $t_{at}$ is an advancing time of the similarity solution,
which is chosen such that $\| \delta P \|_{tl}$ has the minimum value
around $t = 1000$.
Specifically, $t_{at} = 8.5$, $8.3$, $8.3$ and $8.2$ 
for the telegraph model with $w = 0.1$,
the telegraph model 
with $w = 1$, the Palm model with $w = 0.1$
and the Palm model with $w = 1$, respectively. 
These values are consistent with the above optimal value of $t_0$
in Eq.~(\ref{eq:Palm_diff}),
resulting from the fitting (Fig.~\ref{fig:fitting}).
As is obvious from the figure, 
$\| \delta P \|_{tl}$ is 
reduced by two or three orders of magnitude compared with 
$\| \delta P \|$.
Moreover, the algebraic decrease is no longer seen 
for both short and long times.

\begin{figure}
\begin{tabular}{cc}
\scalebox{0.6}{
\includegraphics{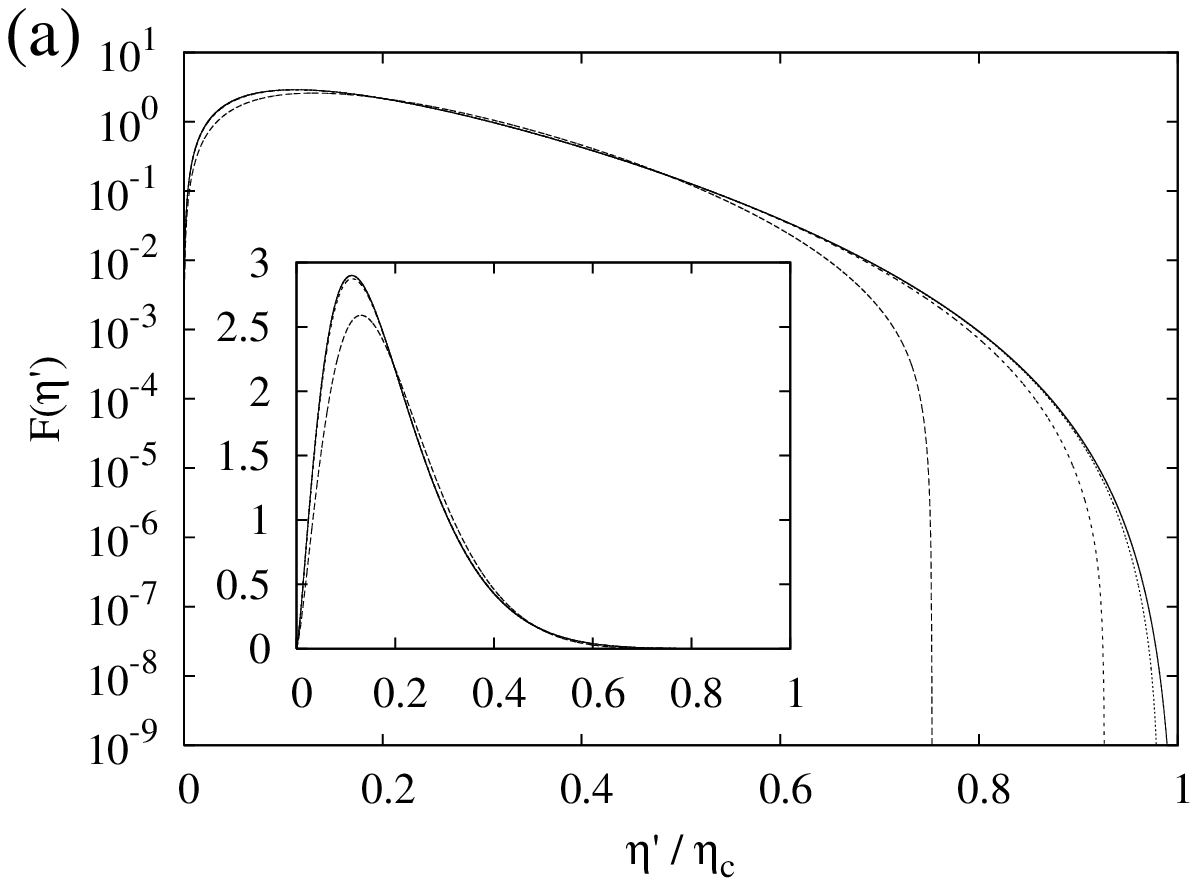}
} &
\scalebox{0.6}{
\includegraphics{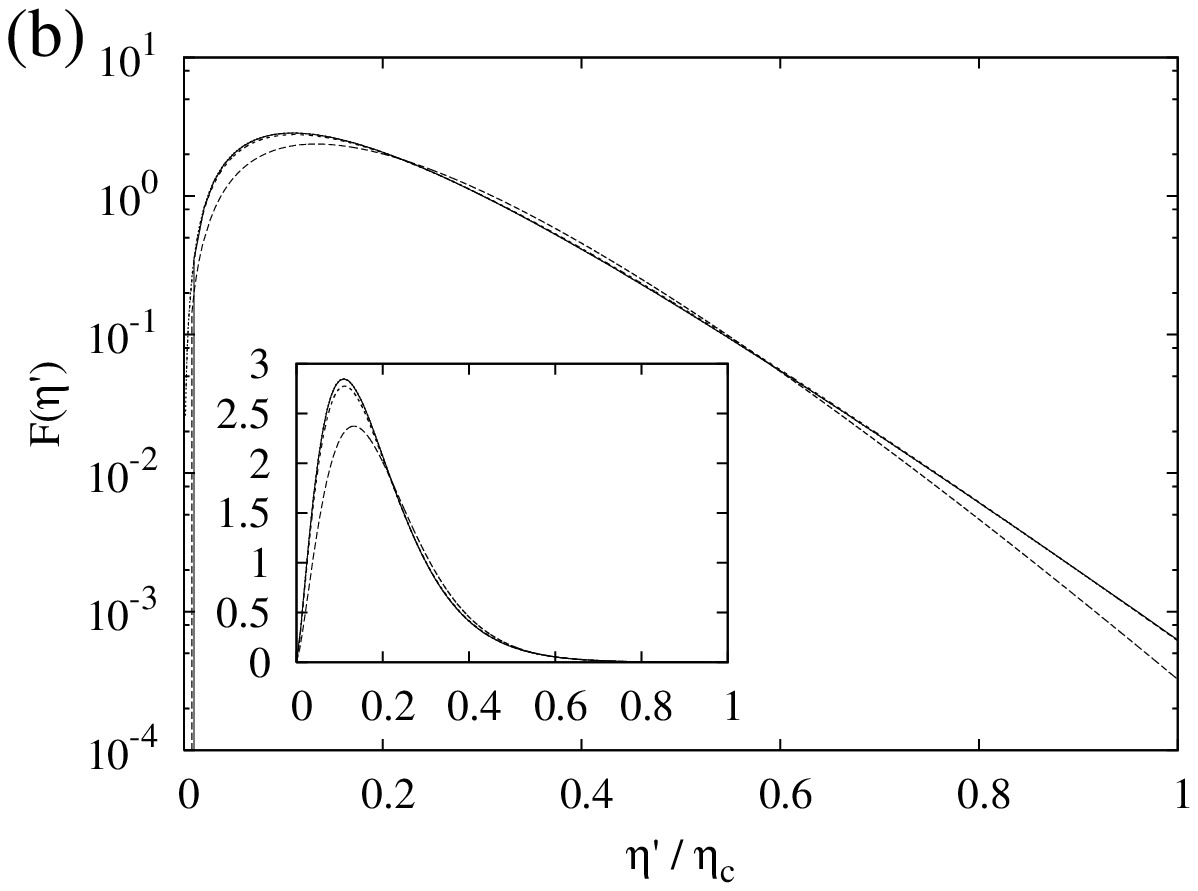}
}
\end{tabular}
\caption{\label{fig:approach_lag}
Same as Fig.~\ref{fig:approach} but for another variable $\eta' = (\lambda r^s) / (t + t_{at})$ with $t_{at} = 8.3$ and $8.2$ for the self-similar telegraph and the Palm model, respectively. Here, the solutions at $t = 1020$ and $4092$ are omitted, since they are almost indistinguishable from the corresponding similarity solution.
}
\end{figure}

In Fig.~\ref{fig:approach_lag}, we show the approach of
the time-integrated solutions to the corresponding similarity solutions
advanced by $t_{at} = 8.3$ 
and $8.2$ for the self-similar telegraph and the Palm model, respectively.
From the comparison between 
Figs.~\ref{fig:approach} and \ref{fig:approach_lag},
it is obvious that the time-integrated solutions
approach the similarity solutions with the time lag much faster than
those without the time lag,
as is expected from Fig.~\ref{fig:diffs}.
At $t = 252$ already, the time-integrated solutions
almost accord with the similarity solutions.
Hence, the approaching time to the similarity solution
becomes shorter than that without the time lag as in Fig.~\ref{fig:approach},
by one order of magnitude.

\subsection{Batchelor scaling}

For times shorter than the 
characteristic time of an eddy of size of the initial separation,
the difference between the mean-square separation
and its initial value is expected to grow
in time as $t^2$,
reflecting the initial ballistic or persistent motion of particle.
This scaling law was first derived by Batchelor
\cite{batchelor50}.
and we refer to it as Batchelor scaling.

\begin{figure}
\scalebox{0.6}{
\includegraphics{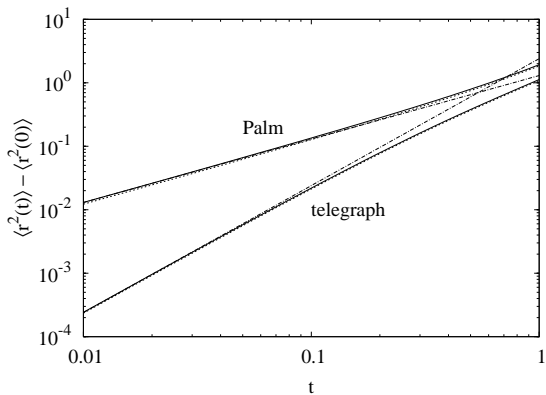}
}
\caption{\label{fig:Batchelor}
Short-time behaviors of the mean-square separations. Here, their initial values, $\langle r^2(0) \rangle = 1 + (1/3 - 2/\pi^2) w^2$, are subtracted. The solid lines are for $w = 1$ and the dashed ones for $w = 0.1$. The straight dot-dashed lines represent $1.3t$ and $2.4t^2$ for the Palm and the self-similar telegraph model, respectively, for reference.  
}
\end{figure}

To see this Batchelor scaling for the present 
time-integrated solutions,
we plot in Fig.~\ref{fig:Batchelor}
$\langle r^2(t) \rangle - \langle r^2(0) \rangle$ versus time
for the above solutions. 
For the self-similar telegraph model
this scaling law is clear.
This result implies that the time derivative of 
the mean-square separation vanishes
at the initial time for the self-similar
telegraph model,
which reminds us of the initial condition
$[\partial P(r,t) / \partial t]_{t=0} = 0$.
However, the scaling law proportional to $t$
is observed for the Palm model.
This arises from 
its incapability
to represent the condition
$[d \langle r^2(t) \rangle / dt]_{t=0} = 0$,
or equivalently 
$[\partial P(r,t) / \partial t]_{t=0} = 0$.
Therefore, the Palm model cannot satisfy
the initial symmetry condition of pair separation
mentioned in Subsec.~\ref{sec:conditions}.
Notice that from Fig.~\ref{fig:Batchelor}
the behaviors for the two different initial widths 
are almost the same, and hence it is concluded that the presence of the ridges 
only in the case of $w = 0.1$ for the telegraph model at short times, 
shown in Figs.~\ref{fig:PDFs}(a) and \ref{fig:PDFs}(c), 
does not affect this scaling law.

In the previous subsection, we showed that
there exists the optimal time of the similarity solution
for minimizing the relaxation time.
This may lead to another scaling law of the mean-square separation.
If we assume that the root mean-square separation
obeys the same temporal evolution
as that of the maximum separation (\ref{eq:separation})
except the coefficient of the time,
we may write the mean-square separation as
\begin{equation}
	\langle r^2(t) \rangle = 
	(\langle r^2(0) \rangle^{s/2} + ct)^{2/s},
\label{eq:mean-square}
\end{equation}
where $c$ is a constant.
This expression was also proposed by Goto and Vassilicos
\cite{goto04:_partic}
using the concept of the doubling time, 
and was adopted in the recent paper
\cite{franzese07}.
Equation~(\ref{eq:mean-square}) can be interpreted 
as the Richardson law with the 
advancing time, $\langle r^2(t) \rangle \propto (t_{at} + t)^{2/s}$,
if we regard $\langle r^2(0) \rangle^{s/2}$ as $ct_{at}$.
This means that the scaling law (\ref{eq:mean-square})
exactly holds for the similarity solution advanced by $t_{at}$.
In fact, Eq.~(\ref{eq:mean-square}) was employed to determine
the Richardson constant in DNS with finite initial separations
 in Ref.~\onlinecite{goto04:_partic}.

\begin{figure}
\begin{tabular}{cc}
\scalebox{0.6}{
\includegraphics{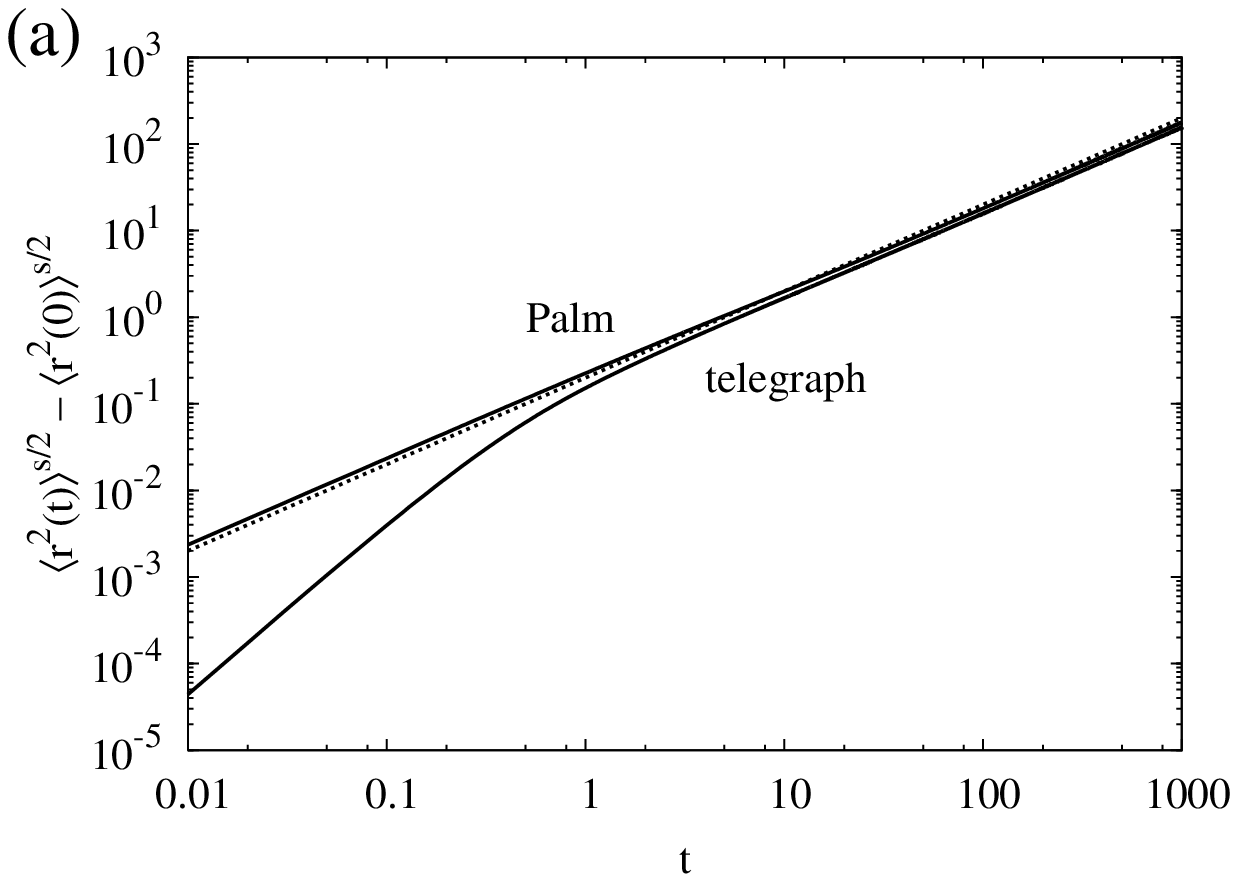}
} &
\scalebox{0.6}{
\includegraphics{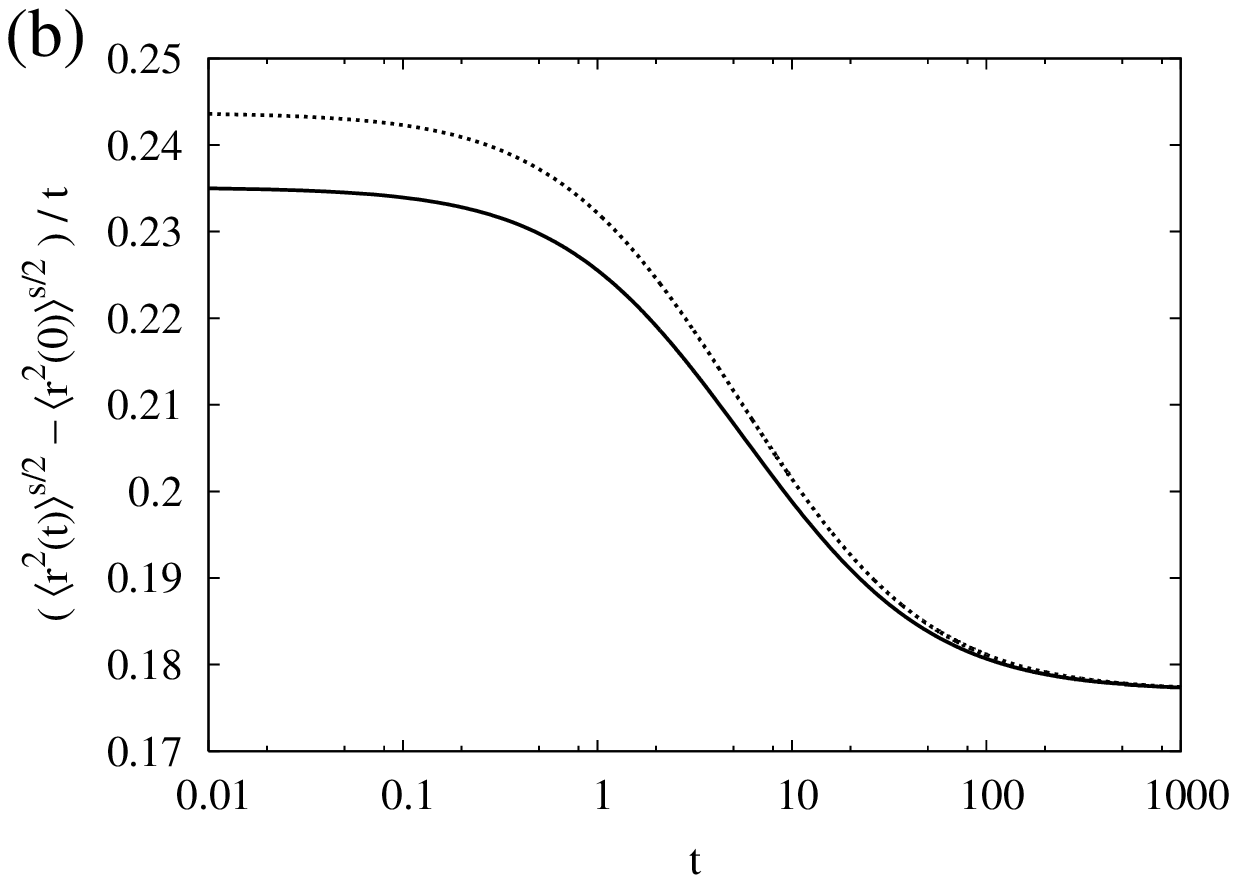}
}
\end{tabular}
\caption{\label{fig:scaling} 
Comparison of the scaling law (\ref{eq:mean-square}) with the
results of the simulations. Panel (a) corresponds to $ct$ and (b) to $c$
in Eq.~(\ref{eq:mean-square}). The straight dashed line in (a)
represents $0.2t$ for reference. $w = 1$ for (a). The solid and the
dashed line in (b) denote the cases of $w = 1$ and $0.1$, respectively,
for the Palm model. 
}
\end{figure}

In Fig.~\ref{fig:scaling}(a), this scaling law
is observed for the Palm model.
Note that this scaling law tells nothing about
the Batchelor scaling.
In fact, as is seen in Fig.~\ref{fig:scaling}(a),
Eq.~(\ref{eq:mean-square}) is not satisfied 
by the self-similar telegraph model,
which satisfies the Batchelor scaling. 

Figure~\ref{fig:scaling}(b) shows the time dependence
of the coefficient $c$ in the right hand side of 
Eq.~(\ref{eq:mean-square}) for the Palm model
with $w = 1$ and $0.1$.
The values of $c$ slightly vary
in the region $1 < t < 10$,
which means that 
Eq.~(\ref{eq:mean-square}) cannot exactly capture 
the behavior of the mean-square separations
of the time-integrated solutions in the whole time regime.
Moreover, the basic assumption $\langle r^2(0) \rangle^{s/2} \simeq ct_{at}$
and the estimated value of the advancing time $t_{at}$ 
in the previous subsection lead to the value of about $0.12$ for $c$ 
for each width of the initial distribution of pair separation.
This value conflicts with that obtained in Fig.~\ref{fig:scaling}(b).
Therefore, the advancing time $t_{at}$ of the similarity solution, 
introduced in the previous subsection, does not have 
a significant meaning for the short time behavior.
The variation of $c$ is smaller for $w = 1$, implying that
Eq.~(\ref{eq:mean-square}) is more appropriate
for the smoother initial PDF,
and becomes less appropriate as the initial PDF gets sharper and
closer to the shape of the delta function.
In the long-time limit this value is related
to the Richardson constant, which is the subject
of the following subsections.

\subsection{Richardson scaling}

\begin{figure}
\scalebox{0.6}{
\includegraphics{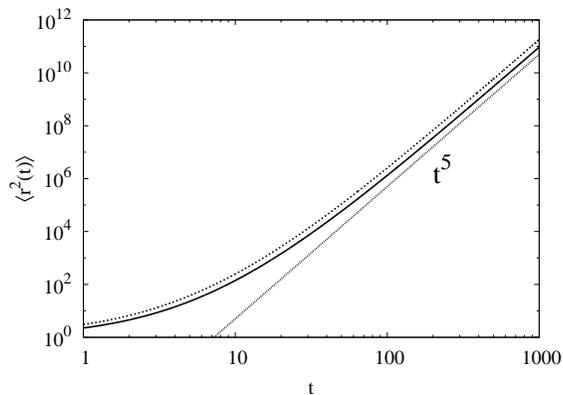}
}
\caption{\label{fig:Richardson}
Long-time behaviors of the mean-square separations. The solid line refers to the self-similar telegraph model and the dashed one the Palm model. $w = 1$. The dotted line is proportional to $t^5$.
}
\end{figure}

For times much longer than the timescale of 
the initial separation
and much shorter than the integral timescale,
the well-known Richardson $t^3$ law 
of the mean-square separation is expected to hold
in 3DNS or 2DIC turbulence 
governed by the Kolmogorov scaling.
In the Bolgiano-Obukhov scaling, however,
the mean-square separation should grow as $t^5$.
This scaling law is clearly seen for the two models
in Fig.~\ref{fig:Richardson} for the long time.

However, now that we previously showed 
the 
agreement of the time-integrated solutions with the similarity solutions
in the long-time limit,
we can use the similarity solutions 
to see that scaling law. 
From the similarity form of the PDF 
$F(\eta) = (t / \lambda)^{1/s} P_s(r,t)$,
it is easy to derive
\begin{equation}
	\langle r^2 \rangle =
	\left(\frac{t}{\lambda} \right)^{2/s}
	\frac{1}{s} \int \eta^{3/s - 1} F(\eta) d\eta.
\label{eq:m.s.s.} 
\end{equation}
Here, the integration on the right hand side
is regarded as a constant for the similarity solutions,
so that we can rewrite Eq.~(\ref{eq:m.s.s.}) as
\begin{equation}
	\langle r^2 \rangle = G \left(
	\frac{t}{\lambda} \right)^{2/s}.
\label{eq:rewritten}
\end{equation}
Thus we can find the $t^5$ dependence of the mean-square 
separation in the Bolgiano-Obukhov scaling
$s = 2/5$.

\subsection{\label{sec:Richardson} Richardson constant}

One of the main interests in turbulent relative dispersion
is the determination of the universal coefficient 
of the Richardson law, namely the Richardson constant
\cite{sawford01:_turbul}.
From Eq.~(\ref{eq:rewritten}), 
$G (C_{\check{A}}/ \lambda)^{2 / s}$ is regarded as the Richardson constant, where $C_{\check{A}}$
is the nondimensional part of 
the dimensional constant $\check{A}$
(see Appendix~\ref{sec:estimate}).

First, we determine the value of $G$.
To do this,
we have only to calculate the integration 
in Eq.~(\ref{eq:m.s.s.})
for the similarity solution.
For the Palm model,
this can be easily carried out using the similarity solution
(\ref{eq:similarity}), and in terms of the gamma function
we write it as
\begin{equation}
	G = s^{4/s} \Gamma
	\left(\frac{2s - \delta + 2}	{s} \right) \bigg/ 
	\Gamma \left(\frac{2s - \delta}{s} \right).
\label{eq:Richardson_Palm}
\end{equation}
Note that this value is independent of the value of $\lambda$.
This is because the Palm equation (\ref{eq:Palm2})
becomes independent of $\lambda$ 
if we use rescaled time
$\check{t} = t / \lambda$.

\begin{figure}
\begin{tabular}{cc}
\scalebox{0.6}{
\includegraphics{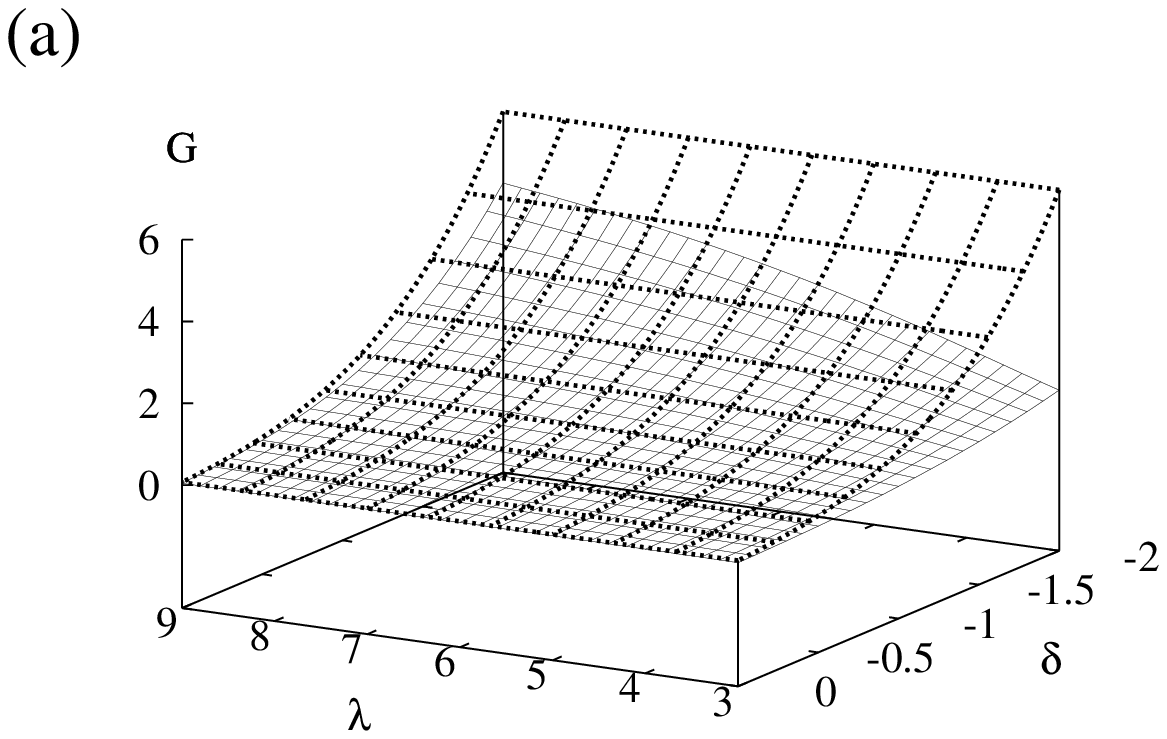}
} &
\scalebox{0.6}{
\includegraphics{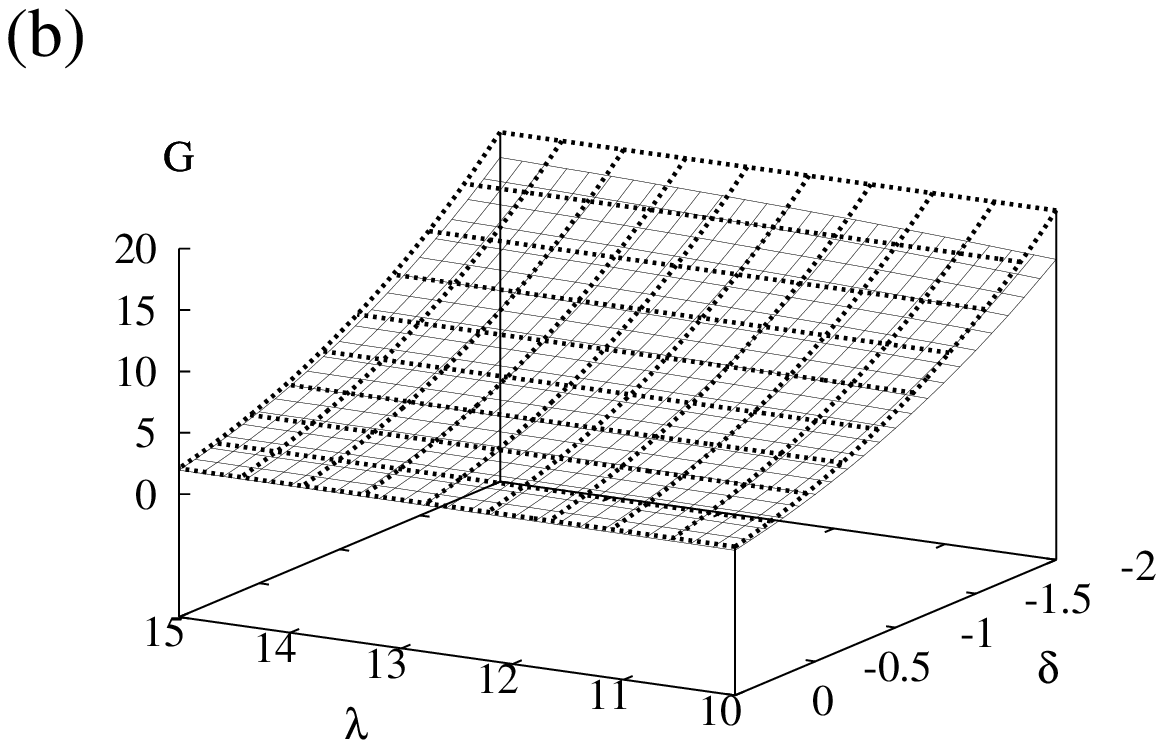}
}
\end{tabular}
\caption{\label{fig:C_R}
The coefficient of the Richardson $t^{2/s}$ law (\ref{eq:rewritten}) for the similarity solutions, $G$, as a function of the parameters $\lambda$ and $\delta$. The upper dashed lines refer to the Palm model, Eq.~(\ref{eq:Richardson_Palm}), and the lower solid lines the self-similar telegraph model. (a): $s = 2 / 5$ for the Bolgiano-Obukhov scaling, (b): $s = 2 / 3$ for the Kolmogorov scaling.
}
\end{figure}

For the self-similar telegraph case, however,
the dependence of $\lambda$ cannot 
be removed by this rescaled time,
and hence $G$ depends on $\lambda$.
Although we do not have the explicit expression of $G$
for this case,
we can evaluate it numerically, and
the evaluated value are shown in Fig.~\ref{fig:C_R}
for various realistic values of $\lambda$ and $\delta$
with those of the Palm 
model.
For $s = 2/5$, $\lambda = 5.2$ and $\delta = -0.77$,
we have $G \cong 0.34$ and $0.66$ 
for the self-similar telegraph
and the Palm model, respectively.

The value of $G$ for the self-similar telegraph case 
accords with that for the Palm case
in the limit of $\lambda \rightarrow \infty$,
since Eq.~(\ref{eq:telegraph2}) with the rescaled time
$\check{t}$ tends to the Palm equation for this limit.
As can be seen from Fig.~\ref{fig:C_R},
$G$ of the Palm model is always larger than
that of the self-similar telegraph.
This means that the inclusion of the effect of
persistent separation for the model 
suppresses the relative dispersion.

Now, we calculate the Richardson constant
for the two models.
The remaining component of the Richardson
constant, $C_{\check{A}}/ \lambda$,
must be determined from the characteristics
of turbulent flows.
This factor is estimated in the way summarized
in the Appendix~\ref{sec:estimate}. 
Note that the value of $\lambda$ is also
required for the self-similar telegraph model, 
since $G$ depends on $\lambda$ in this case.
For our previous DNS of 
2DFC turbulence,
we have estimated the Richardson constant
as $0.030$ and $0.058$ for the self-similar telegraph 
and the Palm model, respectively.
These values are smaller than those estimated of the Richardson constant
for the 2DIC case, also shown in Table~\ref{tab:table3},
by two orders of magnitude.
This might be one of the peculiarities of relative dispersion
in 2DFC turbulence.
From the upper two cases in Table~\ref{tab:table3},
the Richardson constants estimated using the Palm model
appear closer to 
the values obtained by their DNSs (the rightmost column). 
However, in their original papers\cite{boffetta02:_statis,goto04:_partic},
additional assumptions were imposed to obtain their values,
shown in the rightmost column of Table~\ref{tab:table3}:
the satisfaction of the Richardson diffusion equation 
for Ref.~\onlinecite{boffetta02:_statis} and the scaling law
of Eq.~(\ref{eq:mean-square}) for Ref.~\onlinecite{goto04:_partic},
both of which do not hold for the telegraph model.
Therefore, their values cannot be used to determine which model
yields a more correct value of the Richardson constant.

\begin{table*}
\caption{\label{tab:table3}
Estimated values of the Richardson constant
and the relevant parameters.
The dashes denote unavailable values from the papers.
Here, the values are estimated by us, 
except for the rightmost column.
For the estimation of $\lambda$ and $\delta$,
see Ref. \onlinecite{ogasawara06:_turbul}. \\
}
\begin{ruledtabular}
\begin{tabular}{cccccccc}
\\
& \multicolumn{4}{c}{parameters}&\multicolumn{3}{c}{Richardson constant from}\\
paper&$s$&$C_{\check{A}}/\lambda$&$\lambda$&$\delta$
&the telegraph&the Palm&the original\\
&&&&
&model (\ref{eq:telegraph})&model (\ref{eq:Palm})&paper
\\
\hline \\
 Boffetta and Sokolov
\cite{boffetta02:_statis}
&$2/3$&$0.72$&$14$&$-1.48$&
$1.2$&$4.4$&$3.8$ \\
 Goto and Vassilicos
\cite{goto04:_partic}
&$2/3$
 &$0.84$&---&$-0.87$&---&$3.9$&6.9\\
 Ogasawara and Toh
\cite{ogasawara06:_turbul}
&$2/5$&$0.62$&$5.2$&$-0.77$&$0.030$
 &$0.058$&---\\
\end{tabular}
\end{ruledtabular}
\label{table:Richardson's_constant}
\end{table*}

\section{Concluding remarks}

We have numerically solved the two different equations
describing the temporal evolution of the PDF of
the separation of a particle pair in the inertial range
of homogeneous and isotropic turbulence.
In the simulations, 
2DFC turbulence case is dealt with, because we have 
the values of the control parameters for
 that case
\cite{ogasawara06:_turbul}.
The time-integrated solutions of both equations have been compared
to characterize the models.
There, the two initial conditions of 
the different widths of the initial PDF are imposed.
However, the difference between the two widths
only affects the short-time behavior of the PDF
for the self-similar telegraph model.

The self-similar telegraph model represents
 the finiteness of the separation
and its persistency 
from the bounds of PDFs or the maximum relative separations,
appeared in Figs.~\ref{fig:PDFs} and \ref{fig:solutions}, 
whose behavior is not seen in the
 diffusion-type counterpart
(the Palm model).
The inclusion of the effect of persistent separation
in the self-similar telegraph model 
would have an advantage in describing
the initial persistent separation,
which was recently found to strongly affect
the relative dispersion for a relatively long time
in experiments \cite{m.06,n.06}.

In the case of 
the self-similar telegraph model,
when the width of the initial condition (\ref{eq:initial}) gets  smaller, 
the ridges 
appear 
at the 
bounds of the PDF
in the short-time regime.
The uniqueness of the relative velocity
is responsible for this, as is also shown in the model of 
Ref.~\onlinecite{sokolov99:_drude} based on the similar approach to ours.
However, in the real turbulent dispersion,
the relative velocity has a distribution,
and therefore these ridges 
cannot be observed in DNS and 
experiments. 
We will incorporate  the distribution of the relative velocity into 
our model in future work.

The long-time behaviors of the 
time-integrated solutions are characterized by
a relaxation process into the corresponding similarity 
solutions.
It is found from the numerical simulations 
that the similarity solutions 
of both equations, (\ref{eq:telegraph2}) and (\ref{eq:Palm2}),
are asymptotic solutions of the time-integrated solutions.
Thus, the realizability of the similarity solutions
was corroborated, aside from the very long approaching time.
We also investigated the decay rate of the root-squared difference
between the time-integrated and the similarity solutions,
$\| \delta P \|$.
The dominant contribution 
to $\| \delta P \|$ is the time lag between the two solutions.
This contribution seems to be independent of both models and widths
of the initial PDF.

If we compare the time-integrated solution with the 
appropriately-advanced similarity solution, 
more rapid approach to 
it can be found.
This is because the similarity solution is located at the origin 
at the initial time, 
while our initial condition (\ref{eq:initial}) allows for 
nonzero initial separations.
This result may give suggestions to the observation
of the actual relaxation process of separation PDF
in experiments or DNSs.

The Batchelor scaling law for the mean-square separation
holds only for the self-similar telegraph model,
while both models satisfy the Richardson scaling law.
This difference for the short-time behavior between the two models 
is due to their capacities to satisfy the condition 
$[d \langle r^2(t) \rangle / dt]_{t=0} = 0$, or
$[\partial P(r,t) / \partial t]_{t=0} = 0$.

On the other hand, 
the behavior of the mean-square separation for the Palm model can be 
well described by the Richardson scaling law with the time lag,
Eq.~(\ref{eq:mean-square}), to some extent.
However, 
as is suggested by the fact that the Batchelor scaling cannot be derived 
from Eq.~(\ref{eq:mean-square}),
this scaling law neglects the effect of initial persistent separation
of a particle pair.

We have also estimated the 
Richardson constant for both models 
making use of the data of DNSs 
to determine the values of the parameters of the models.
The Richardson constant for the self-similar 
telegraph model is generally smaller than that for the Palm model.
Furthermore, both models predict smaller values for the 2DFC case 
than those predicted for the 2DIC case, 
by a factor of $10^{-2}$.
This result should be confirmed in future DNS of 2DFC turbulence. 

Although we have 
dealt with the 2DFC turbulence,
we have not yet observed the clear Richardson $t^5$ law
in that case \cite{ogasawara06:_turbul}.
We 
will attempt to 
achieve the Richardson $t^5$
law as well as the separation PDF  in future work 
 to determine which model 
is better for describing the relative dispersion.

\begin{acknowledgments}
  This work was supported by the Grant-in-Aid for the 21st Century COE
  ``Center for Diversity and Universality in Physics'' from the Ministry
 of Education, Culture, Sports, Science and Technology (MEXT) of Japan.
 Some of the numerical computations in this work were carried out on NEC
  SX-8 at the Yukawa Institute Computer Facility.
 K.K. thanks T. Matsumoto for 
 suggestions on the numerical method and for invaluable 
 discussions. 
  S.T. was partly supported by the Grant-in-Aid for Scientific Research
 (C) (18540373) from Japan Society for the Promotion of Science.
\end{acknowledgments}

\appendix

\section{\label{sec:decrement}
Loss of the total probability for self-similar telegraph model}

\begin{figure}
\begin{tabular}{cc}
\scalebox{0.6}{
\includegraphics{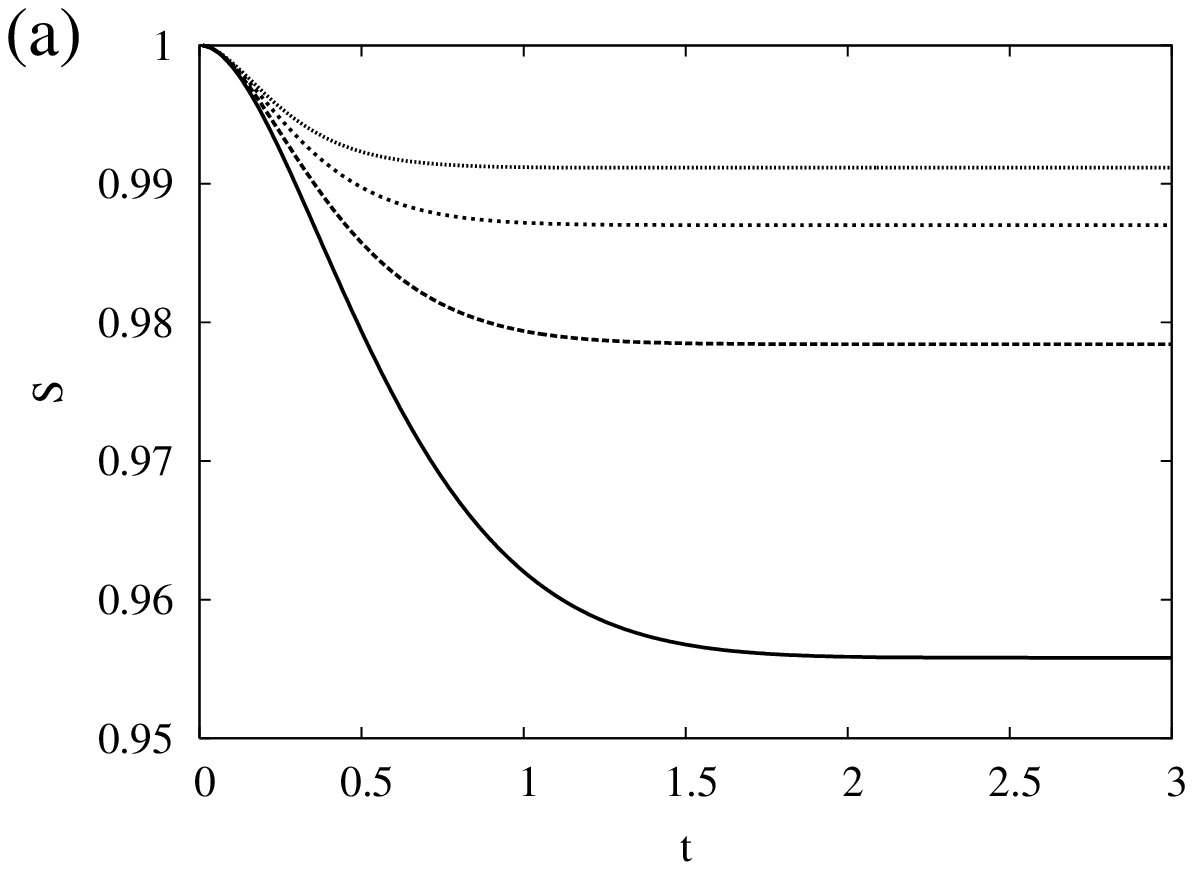}
} &
\scalebox{0.6}{
\includegraphics{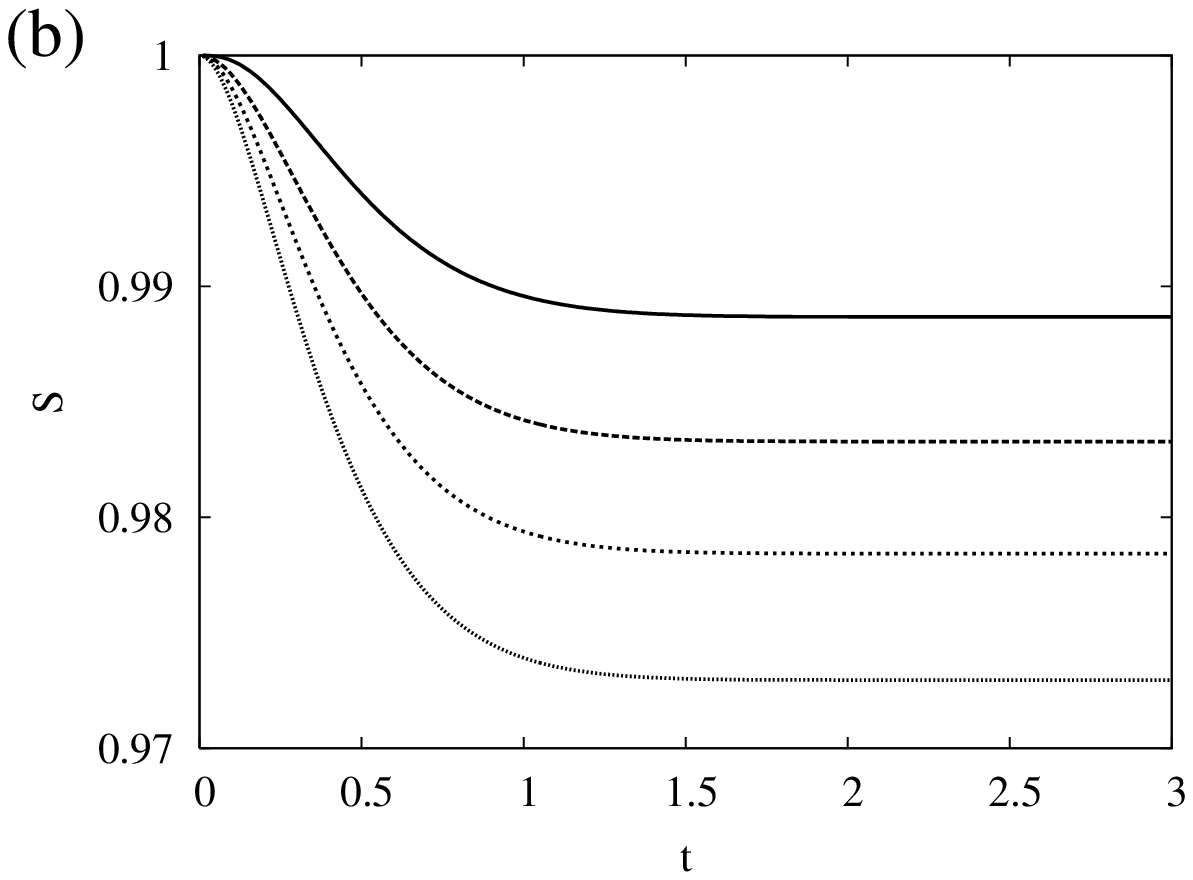}
}
\end{tabular}
\caption{\label{fig:total} 
Temporal evolution of the total probability, $S$, for various values of the parameters $\lambda$ and $\delta$. (a): $\delta = -0.77$ and $\lambda = 3.5$, $5.2$, $6.8$ and $8.3$ from bottom to top. (b): $\lambda = 5.2$ and $\delta = -1.2$, $-0.77$, $-0.4$ and $0$ from bottom to top. $w = 1$ for both. The values remain constant after $t = 2$.
}
\end{figure}

Although for the Palm model
the total probability $S = \int P(r,t) dr$
is conserved,
the self-similar telegraph model does not assure
the conservation of $S$.
Thus we calculated $S$ using the 
time-integrated solution 
and showed its temporal evolution
in Fig.~\ref{fig:total},
varying the values of the parameters 
$\lambda$ and $\delta$.
We can easily see from this figure
that $S$ monotonically decreases for $t < 2$
and is conserved afterward.
The variation of the value of $w$
does not change these graphs, even quantitatively.
Therefore, we may regard these results as universal,
independent of the initial condition.
Note that the loss of $S$ is not 
the result of errors in accuracy of 
the numerical scheme used,
since finer grids and time steps
make the same figures.
As is mentioned in our previous paper,
the similarity solution fulfills the
probability conservation.
However, we cannot conclude from this that the 
time-integrated solution already has the similarity form
for $t > 2$.

\begin{figure}
\begin{tabular}{cc}
\scalebox{0.6}{
\includegraphics{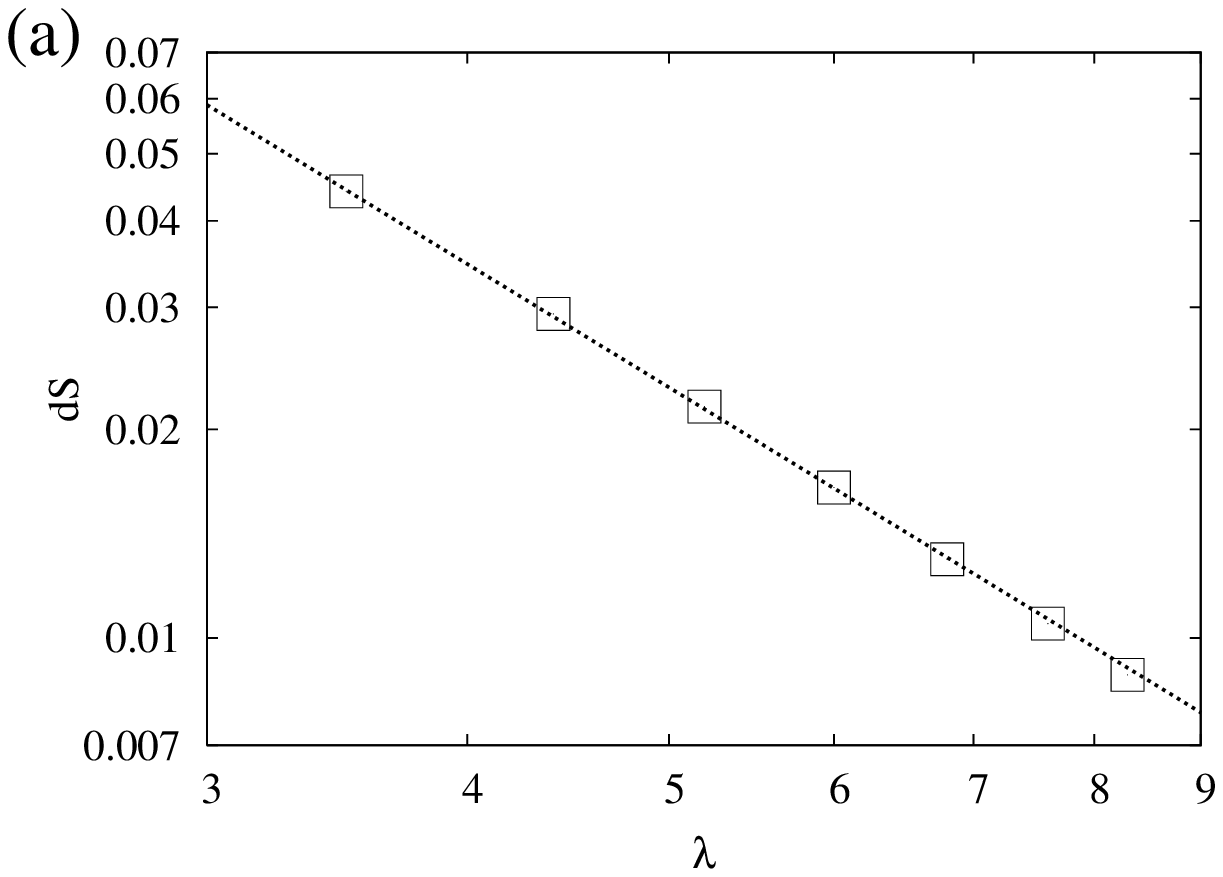}
} &
\scalebox{0.6}{
\includegraphics{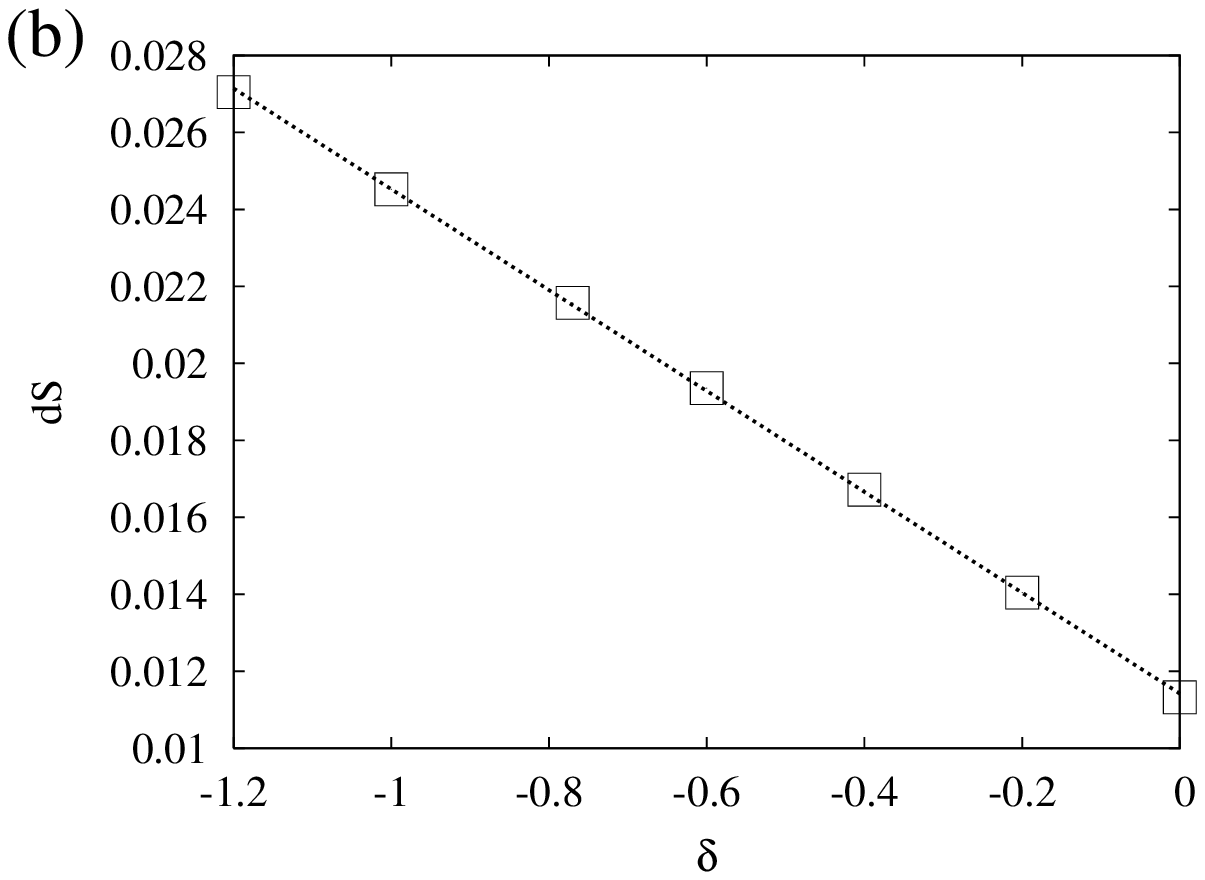}
}
\end{tabular}
\caption{\label{fig:decrement}
Dependence on the parameters $\lambda$ and $\delta$ of the loss of the total probability, $dS$, at $t = 2$. (a): $\delta = -0.77$, (b): $\lambda = 5.2$. The lines are fits to the data and represent $0.44\lambda^{-1.8}$ and $-0.013\delta + 0.011$ for (a) and (b), respectively. $w = 1$ for both.}
\end{figure}

We also plotted the dependence on the parameters
$\lambda$ and $\delta$
of the loss of the total probability,
$dS = 1 - S$, at $t = 2$
in Fig.~\ref{fig:decrement}.
There the scaling laws $0.44\lambda^{-1.8}$ and
$-0.013\delta + 0.011$ are seen.
Its decrease with increasing $\lambda$
is obvious from the fact that the self-similar
telegraph equation~(\ref{eq:telegraph2})
tends to the Palm equation~(\ref{eq:Palm2})
as $\lambda \rightarrow \infty$,
keeping the rescaled time $\check{t}$
appeared in Subsec.~\ref{sec:Richardson} unchanged.

It is not difficult to make the self-similar telegraph model to
conserve the total probability by reformulating it in a conserved form.
The results will be reported in future papers.

\section{\label{sec:estimate}
estimate of $C_{\check{A}}/ \lambda$}

The dimensional constant $\check{A}$ can be decomposed 
into the nondimensional and dimensional parts as
$C_{\check{A}} \varepsilon^{1/3}$
and $C_{\check{A}} \varepsilon_\theta^{1/5} 
(\alpha g)^{2/5}$ for the Kolmogorov and 
Bolgiano-Obukhov scaling, respectively.
Here, the nondimensional part is denoted by $C_{\check{A}}$.
The dimensional constants
$\varepsilon$, $\varepsilon_\theta$, $\alpha$
and $g$ are the energy dissipation rate,
the entropy dissipation rate,
the thermal expansion coefficient and
the gravitational acceleration, respectively. 

In order to determine the value of 
$C_{\check{A}}/ \lambda$,
we use the exit time statistics.
The exit time, $T_E(r;\rho)$, is the time it takes
for the separation of the particle pair
to reach the threshold of $\rho r$ 
from that of 
$r$.
Assuming that the PDF of the pair separation
obeys Eq.~(\ref{eq:Palm}), 
the following expression of the mean exit time,
\begin{equation}
	\langle T_E(r;\rho) \rangle =
	\frac{\lambda}{\check{A}} \frac{1}{s(2s - \delta)}
	(\rho^s - 1) r^s,
\label{eq:exit-time}
\end{equation}
is derived 
\cite{ogasawara06:_turbul}
in the same manner as Ref.~\onlinecite{boffetta02:_statis}. 
The proportionality of the mean exit time to
$\check{A}^{-1} (\rho^s - 1) r^s$ is also understood
from the scaling law for the characteristic time,
Eq.~(\ref{eq:time}).
If we set the values of  $s$, $\delta$ and $\rho$ and estimate that
of the proportionality coefficient of $r^s$ in the
right hand side of Eq.~(\ref{eq:exit-time})
from the data of DNSs or experiments,
then the value of $\check{A} / \lambda$
can be calculated.
Thus, we obtain the value of 
$C_{\check{A}}/ \lambda$
dividing $\check{A} / \lambda$ 
by the dimensional part of $\check{A}$,
corresponding to the scaling of the DNS or experiment.


\end{document}